\documentclass[sn-mathphys,Numbered]{sn-jnl}

\usepackage{mathtools}
\usepackage{graphicx}%
\usepackage{multirow}%
\usepackage{amsmath,amssymb,amsfonts}%
\usepackage{amsthm}%
\usepackage{mathrsfs}%
\usepackage{xcolor}%
\usepackage{textcomp}%
\usepackage{manyfoot}%
\usepackage{booktabs}%
\usepackage{algorithm}%
\usepackage{algorithmicx}%
\usepackage{algpseudocode}%
\usepackage{listings}%
\usepackage{soul}%

\DeclareUnicodeCharacter{00A0}{~}

\begin{document}
\title[Probing quantum correlations in non-degenerate hyper-Raman process]{Probing quantum correlations in non-degenerate hyper-Raman process}

\author{
    Moumita Das\(^1\), Biswajit Sen\(^2,\footnote{Corresponding author: bsen75@yahoo.co.in}\), Ankur Sensharma\(^3\), Kishore  Thapliyal\(^4\), Anirban Pathak\(^5\)\\
    \(^1\)Department of Physics, Malda College, Malda-732101, WB, India \\
    \(^2\)Institute of  Physics, Vidyasagar Teachers' Training College, Midnapur-721 101, WB, India \\
    \(^3\)Department of Physics, University of Gour Banga, Malda-732101, WB, India\\
    \(^4\)Joint Laboratory of Optics, Faculty of Science, Palacky University, Czech Republic, 17. listopadu 12, 779 00, Olomouc, Czech Republic\\
    \(^5\)Jaypee Institute of Information Technology, A-10, Sector-62, Noida UP-201309, India
}


\abstract{Possibilities of observing single mode and
intermodal quantum correlations (e.g., antibunching, steering and
entanglement) are studied for a probed-hyper-Raman system with specific
attention on the impact of a probe on the single and multi-mode quantum
correlations generated in a hyper-Raman active system. The physical
system studied here considers that the probe interacts continuously
with the non-degenerate pump modes in the hyper-Raman active system
via a nonlinear coupling. The investigation has revealed that quantum
correlations in the Raman systems can be controlled using the probe.
Further, it is observed that the quantum steering between the pump and
anti-Stokes modes can be influenced significantly by  controlling
the interaction between the system and the probe. Unlike steering,
probe could neither deteriorate the nonclassical correlations, namely
intermodal entanglement and photon antibunching, nor induce them.
Though the witness of the corresponding nonclassical effect depends
on the initial state of the probe as well as the coupling strength.}




\maketitle

\section{Introduction}\label{sec1}
Recently, quantum correlation has generated a great deal of
attention with the progress of quantum technologies \citep{QT} and
quantum information science \citep{QT-2}. Specifically, the facts
that quantum correlations can be exploited to perform tasks having
no classical analogue (e.g., teleportation \citep{Bennet1993}, dense
coding \citep{densecoding} and remote state preparation \citep{RSP}),
to provide device independent schemes for cryptography that can provide
unconditional security \citep{DI-QKD}, and to generate self-testable
quantum random number generators \citep{QRNG}, have generated a wide
spread interest on quantum correlations. Further, a class of correlations
are related to the nonclassical states--quantum states having no
classical analogue; and to obtain quantum advantage in any computing
or communication scenario, one would require nonclassicality. Keeping
that in mind, in what follows, we restrict ourselves to the study
of a set of quantum correlations that are relevant to quantum technologies.
In fact, quantum correlations have many facets. Among them quantum
steering, entanglement, nonlocality and photon antibunching are the
important phenomena which illustrate the presence of quantum correlations.
Each of these phenomena are appealing because of the great practical
applications associated with them. Keeping the above in mind, in this
paper, we aim to study the possibilities of observing these phenomena
in the probed-hyper-Raman system.

The rationale behind the selection of the probed-hyper-Raman
system as the test bed for the present study on quantum correlation
underlies in the facts that Raman and hyper-Raman processes are studied
in depth in the works reported by some of the present authors \citep{bsen-subshot,bsen-mandal-hyper,off-res-Raman,off-res-RamanPh,KT-PRA,giri1,giri2}
and other groups \citep{perina-book,Raman-chip,Raman-chip-rev,Raman-Q,Raman-Q1},
but the observed quantum correlations can be impacted by probing and
the earlier studied systems can be obtained as the special cases of
the probed-hyper-Raman system studied here. In this work, our motivation
is to evaluate the effect of a probe on the single and multi-mode
quantum correlations generated in the hyper-Raman active system qualitatively.
Here, the probe interacts with the non-degenerate pump modes in the
hyper-Raman active system via a nonlinear coupling. This continuous
interaction is studied as quantum Zeno or anti-Zeno effect when it
suppresses or enhances the dynamics of the system (\citep{das,das2}
and references therein). This impact of the presence of the probe
on the system dynamics is quantified in terms of change in the average
photon/phonon numbers due to the external probe \citep{das,das2,NZeno}.
The role of this probe on the generation of nonclassicality in a nonlinear
system is still unexplored. For this study, we focus on a quantum
system exhibiting quantum Zeno and anti-Zeno effect \citep{das} to
witness quantum correlations generated in this system.

The rest of the paper is organized as follows. In Section \ref{sec:physical-system},
the physical system of our interest is described as a codirectional
asymmetric nonlinear optical coupler composed of a probe and a system,
which is a nonlinear waveguide operating under hyper-Raman process.
The possibility of observing quantum steering, intermodal entanglement
and antibunching in the system of interest is investigated in Section
\ref{sec:Nonclassicality}. Finally, the findings are summarized and
the article is concluded in Section \ref{sec:Conclusion}.

\begin{figure}
\begin{centering}
{  \includegraphics{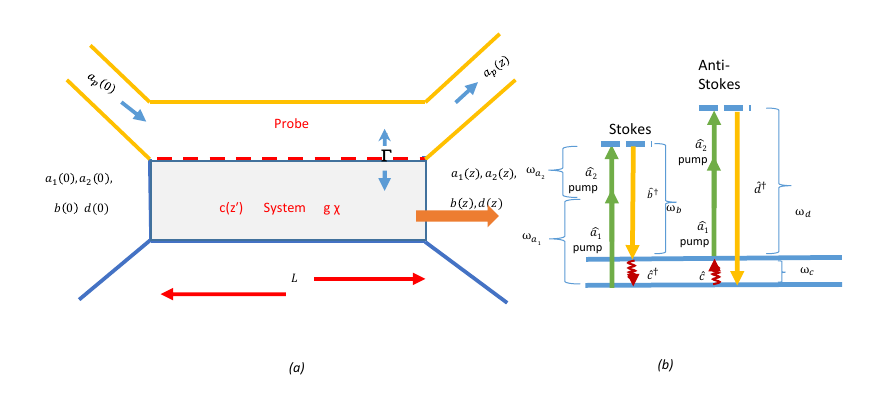}}{ }
\par\end{centering}
{  \caption{\label{fig:physical-system}(Color online) (a) Schematic diagram of an asymmetric nonlinear optical coupler of length $L$ composed
a system waveguide, operating under non-degenerate hyper-Raman process,
interacting with a probe (linear waveguide). (b) Schematic diagram
of the non-degenerate hyper-Raman Process. All the optical (namely,
probe $a_{p}$, non-degenerate hyper-Raman pump $a_{1},\,a_{2}$,
Stokes $b$, and anti-Stokes $d$ and phonon $c)$ modes
with three coupling coefficients $g$, $\chi$, and $\Gamma$
are indicated.}
}{ }
\end{figure}

\section{Physical System \label{sec:physical-system}}

The physical system of our interest is a codirectional asymmetric
nonlinear optical coupler composed of a probe and a system, which
is a nonlinear waveguide operating under hyper-Raman process (see
Fig. \ref{fig:physical-system} for a schematic diagram) and only
the non-degenerate pump modes of the system waveguide interact with the
probe, which is a linear waveguide. The momentum operator of the entire
physical system comprising of the probe and the system is given by 
\begin{equation}
\begin{array}{lcl}
G & = & k_{p}a_{p}^{\dagger}a_{p}+k_{a_{1}}a_{1}^{\dagger}a_{1}+k_{a_{2}}a_{2}^{\dagger}a_{2}+k_{b}b^{\dagger}b+k_{c}c^{\dagger}c+k_{d}\\
 & \times & d^{\dagger}d+\left(ga_{1}a_{2}b^{\dagger}c^{\dagger}+\chi a_{1}a_{2}cd^{\dagger}+\Gamma a_{p}a_{1}^{\dagger}a_{2}^{\dagger}+{\rm H.c.}\right),
\end{array}\label{eq:ham}
\end{equation}
where H.c. denotes the Hermitian conjugate and $\hbar=1$ (the same
convention is followed throughout the work). The annihilation (creation)
operators $a_{p}\,(a_{p}^{\dagger}),\,a_{i}\,(a_{i}^{\dagger}),\,b\,(b^{\dagger}),\,c\,\left(c^{\dagger}\right)$,
and $d\,(d^{\dagger})$ correspond to the probe pump (indexed by subscript
$p)$, pump mode (indexed by subscript $i=1,2$), Stokes mode, vibration
(phonon) mode and anti-Stokes mode, respectively, of entire system.
All the field operator comply to the standard bosonic commutation
relation $\left[A_{i},A_{j}^{\dagger}\right]=\delta_{ij}$. The wave vectors
corresponding to the probe, Pump-1, Pump-2, Stokes, phonon and anti-Stokes
modes are denoted by $k_{p},$ $k_{a_{1}}$, $k_{a_{2}},$ $k_{b}$,
$k_{c}$, and $k_{d}$, respectively. The Stokes and anti-Stokes coupling
constants are described by parameters $g$ and $\chi$, respectively.
Further, $\Gamma$ denotes the interaction constant between the probe
and the two pump modes present in the hyper-Raman process in the system.

We obtain the Heisenberg's equations of motion for the momentum
operator (\ref{eq:ham}) of the system of interest, which gives us
a set of six coupled differential equations for the bosonic operators
in the momentum operator. These are as follows: 
\begin{equation}
\begin{array}{lcl}
\overset{.}{a}_{p}\left(z\right) & = & i\left({k}_{p}a_{p}+\Gamma a_{1}a_{2}\right),\\
\overset{.}{a}{_{1}}\left(z\right) & = & i\left({k}_{a_{1}}a_{1}+ga_{2}^{\dagger}bc+\chi a_{2}^{\dagger}c^{\dagger}d+\Gamma a_{p}a_{2}^{\dagger}\right),\\
\overset{.}{a}{_{2}}\left(z\right) & = & i\left({k}_{a_{2}}a_{2}+ga_{1}^{\dagger}bc+\chi a_{1}^{\dagger}c^{\dagger}d+\Gamma a_{p}a_{1}^{\dagger}\right),\\
\overset{.}{b}\left(z\right) & = & i\left({k}_{b}b+ga_{1}a_{2}c^{\dagger}\right),\\
\overset{.}{c}\left(z\right) & = & i\left({k}_{c}c+ga_{1}a_{2}b^{\dagger}+\chi a_{1}^{\dagger}a_{2}^{\dagger}d\right),\\
\overset{.}{d}\left(z\right) & = & i\left({k}_{d}d+\chi a_{1}a_{2}c\right).
\end{array}\label{eq:equationofmotion}
\end{equation}

This set of coupled nonlinear equations is not exactly solvable in
the closed analytical form. Therefore, we obtain the spatial evolution
of all the operators using Sen-Mandal perturbative technique assuming
dimensionless quantities $gz$, $\chi z$, and $\Gamma z$ are small
compared to unity. All the mathematical details are in
Appendix A, where we have reported the spatial evolution
corresponding to all the field operators for probe, pump, Stokes,
anti-Stokes, and phonon modes using the Sen-Mandal technique. Further,
using the boundary condition, we obtain the decoupled differential
equations, from where we finally got the analytical solutions. These
solutions are exploited to investigate steering, intermodal entanglement,
and photon antibunching for different modes of the physical system
of our interest. It is also crucial to remember that since there
is only one passage configuration \citep{thun2002zeno-raman} in the
degenerate hyper-Raman system and damping is supposed to reduce the
quantum effect, we can ignore the losses due to damping.

Similar asymmetric nonlinear optical couplers were found useful
for generating quantum correlations \citep{kishore2014co-coupler},
where Sen-Mandal perturbative technique was shown advantageous over
some other approximation methods. In the next section, we will show
the feasibility of observing quantum steering and intermodal entanglement
for various coupled modes. Further, we will provide the features of
photon antibunching for the pump mode $a_{i}(i\in\left\{ 1,2\right\} ).$

\section{Nonclassical Properties of Bosonic Modes \label{sec:Nonclassicality}}

To study the nonclassical characteristics of the different
bosonic modes in the degenerate hyper-Raman optical coupler, i.e.,
photon antibunching, quantum intermodal entanglement and steering,
we use the dynamical solution of Eq. (\ref{eq:equationofmotion}).
We consider the initial composite coherent state $\left|\psi(0)\right\rangle $
as the product of the initial coherent states of the probe, pump,
Stokes, phonon, and anti-Stokes modes as $\left|\alpha\right\rangle $,
$\left|\alpha_{j}\right\rangle $, $\left|\beta\right\rangle $, $\left|\gamma\right\rangle $,
and $\left|\delta\right\rangle $, respectively. Hence, the initial
state is considered to be 
\begin{equation}
\begin{array}{lcl}
\left|\psi(0)\right\rangle  & = & \left|\alpha\right\rangle \otimes\left|\alpha_{1}\right\rangle \otimes\left|\alpha_{2}\right\rangle \otimes\left|\beta\right\rangle \otimes\left|\gamma\right\rangle \otimes\left|\delta\right\rangle ,\end{array}\label{eq:inSt}
\end{equation}
and the field operator $a_{p}$ operating on the initial state gives
\begin{equation}
\begin{array}{lcl}
a_{p}(0)\left|\psi(0)\right\rangle  & =\alpha & \left|\alpha\right\rangle \otimes\left|\alpha_{1}\right\rangle \otimes\left|\alpha_{2}\right\rangle \otimes\left|\beta\right\rangle \otimes\left|\gamma\right\rangle \otimes\left|\delta\right\rangle ,\end{array}\label{eq:ann}
\end{equation}
where $\begin{array}{ccc}
\alpha & = & \left|\alpha\right|e^{i\varphi_{p}}\end{array}$ is the complex eigenvalue with $\left|\alpha\right|^{2}$ initial
average number of photons with phase angle $\varphi_{p}$ in the coherent
state $\left|\alpha\right\rangle .$ In the similar manner, coherent
state parameters for all the optical and phonon modes involved in
the hyper-Raman process can be defined as $\Lambda_{j}=\left|\Lambda_{j}\right|e^{i\varphi_{j}}$
for complex amplitudes $\Lambda_{j}:\Lambda\in\left\{ \alpha,\beta,\gamma,\delta\right\} $
and the corresponding phase angles $\varphi_{j}$ with $j\in\left\{ 1,2,b,c,d\right\} $
for the non-degenerate Pump-1, Pump-2, Stokes, phonon, and anti-Stokes
modes, respectively.

Generation of multipartite entangled states in Raman and hyper-Raman
processes is studied in the past \citep{off-res-Raman,off-res-RamanPh,KT-PRA,giri1,giri2}.
The existence of multipartite entanglement signifies mixed states
in the different combinations of two-mode reduced states. Further,
pure state entanglement corresponds to nonlocal correlations \citep{gisin,popescu},
unlike mixed states where steering and Bell nonlocality are stronger
correlations than entanglement. Therefore, we study quantum correlations,
namely steering and entanglement, in bipartite states and autocorrelation
in the optical single modes as antibunching.

\subsection{Quantum Steering}

Quantum steering or simply steering is a concept within quantum
mechanics that explains a type of asymmetric quantum correlation between
two modes/particles that are entangled. Quantum steering describes
the impact of the local projective measurement performed by an observer
on the reduced state of another mode/particle. In other words, the
measurement choice made by one observer can alter the quantum state
of the other particle even if they are separated in space. Quantum
steering plays an important role in quantum communication, quantum
information processing, quantum cryptography, and essential tests of
quantum theory. For instance, quantum steering is a useful resource
for one-way device independent quantum cryptography \citep{1way-QKD}.
In our present work, we use the criterion for the steering \citep{steeringcriteria-1,steeringcriteria-2}
as

\begin{equation}
\begin{array}{lcl}
S_{i\rightarrow j} & = & E_{ij}+\frac{\left\langle i^{\dagger}i\right\rangle }{2}<0,\end{array}\label{eq:steering-criteria}
\end{equation}
where, $i,$ $j$ are any two arbitrary bosonic operators and $\begin{array}{lcl}
E_{ij} & = & \left\langle i^{\dagger}ij^{\dagger}j\right\rangle -\left|\left\langle ij^{\dagger}\right\rangle \right|^{2}\end{array}$ is the entanglement criteria (HZ-1) we will use in the subsection.
If $S_{i\rightarrow j}$ goes below zero, the combined state shows
steering. 

Quantum steering is a unique quantum nonlocal correlation as
it is asymmetric unlike quantum entanglement and Bell nonlocality.
Therefore, mode $i$ may steer mode $j$ without mode $j$ steering
mode $i$ if $S_{i\rightarrow j}<0$ but $S_{j\rightarrow i}>0$. The expressions of all possible intermodal nonclassicality witness for steering are given in Appendix B. Notice that in probed-non-degenerate hyper-Raman system we observe that nonclassical features corresponding to both the pump modes are identical. Therefore, in what follows, we drop the subscript from $a_j$ to mention $a$ for both the pump modes.
Here, we observed that both the pump modes can steer the anti-Stokes
mode without the presence of the probe mode (as shown in the blue
smooth line in Fig. 2(a)). This steering can be observed after a small
rescaled interaction length as $S_{a\rightarrow d}$ becomes negative
after $gz\gtrapprox0.0165$. However, the negativity of the steering
parameter can be regulated by the presence of the probe via controlling
various interaction parameters between the system and the probe. Specifically,
the probe mode initially in the vacuum provides the best scenario
for generating a steered state (as shown in the red dashed line in Fig.
2(a)). However, this advantage of the probe disappears with increasing
probe pump intensity as the rescaled interaction length required for
observing the steering is larger (as shown in the cyan dot-dashed
line in Fig. 2(a)). Controlling the phase parameter of the initial
coherent state in the probe mode allows us to counter this effect
to harness the potential of probing steering in pump-anti-Stokes mode
(as shown in the magenta dotted line in Fig. 2(a)). Specifically,
the negative values of $S_{a\rightarrow d}$ can be achieved for the
smallest rescaled interaction length among all the cases discussed
here.

\begin{figure}[t]
    \centering
    \includegraphics[scale=0.5]{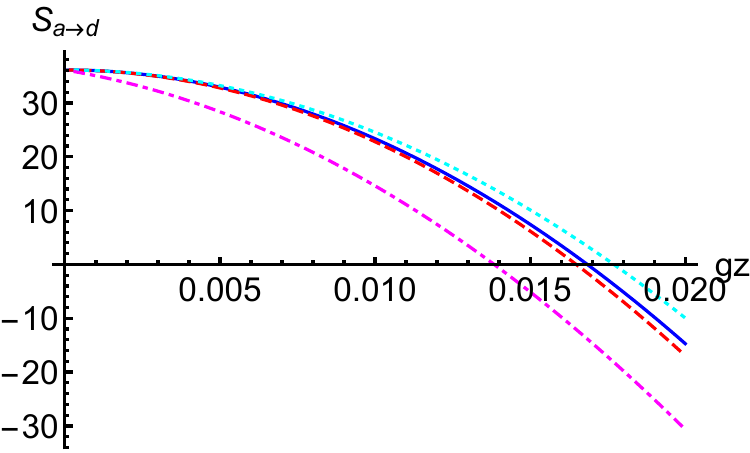} 
    \includegraphics[scale=0.5]{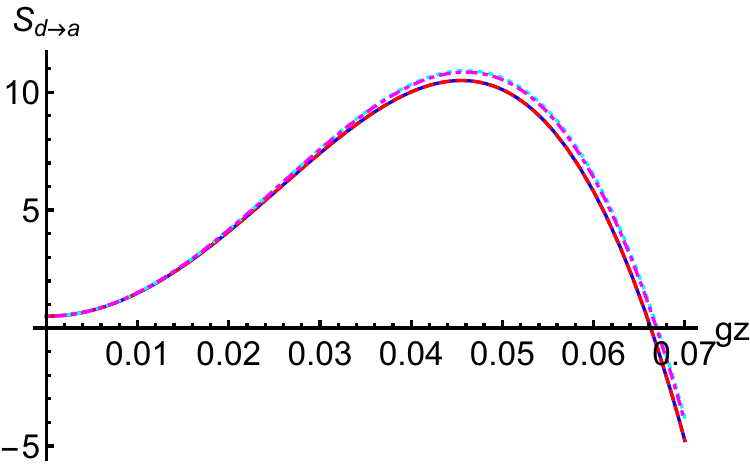}
    \centerline{ \hspace{2mm} (a) \hspace{.45\hsize} (b) }\\
    
    \includegraphics[scale=0.55]{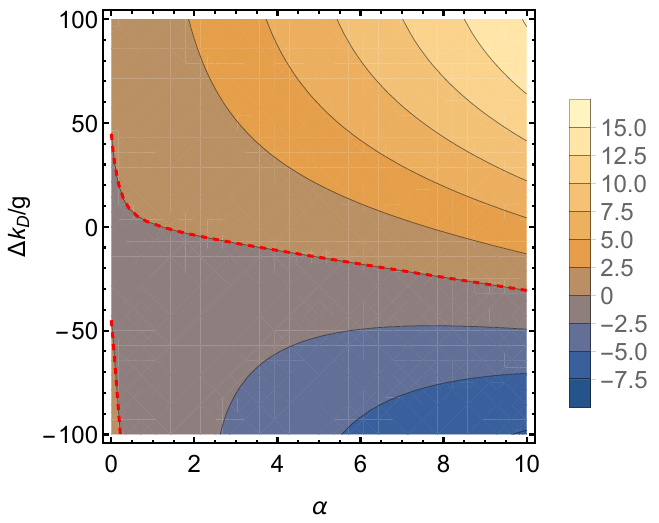} 
    \includegraphics[scale=0.55]{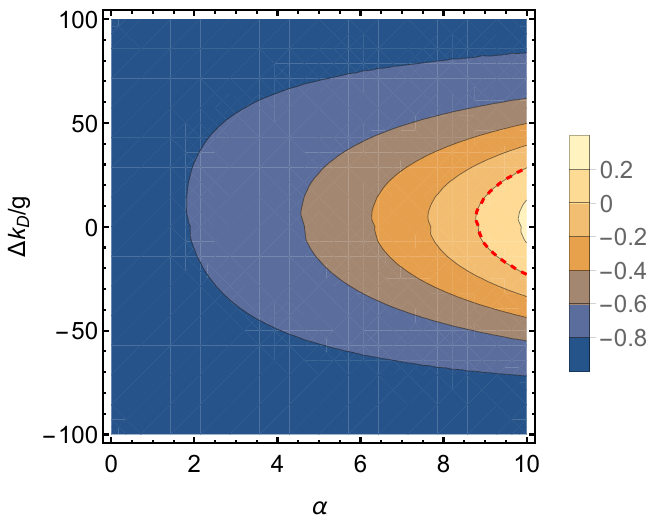}
    \centerline{ \hspace{2mm} (c) \hspace{.45\hsize} (d) }\\
    
    \caption{\protect\label{fig:steer} (Color online)
    (a) Pump steering anti-Stokes and (b) anti-Stokes steering pump for
    $\Gamma=0$ shown in smooth (blue), $\Gamma=1.5g,\alpha=0$ dashed
    (red), $\Gamma=1.5g,\alpha=9$ dot-dashed (cyan), and $\Gamma=1.5g,\alpha=9\exp\left(i\pi\right)$
    dotted (magenta) lines, respectively. We have chosen here the coupling
    coefficients $\chi=1.2g$ with initial coherent state amplitudes for
    all modes $\alpha_{1}=8.5,\alpha_{2}=8.3,\beta=7,\gamma=0.01,\delta=-1$.
    We have also used $\Delta{k}_{S}=-10g,\Delta{k}_{A}=19g,\Delta{k}_{D}=9$.
    Steering parameters (c) $S_{a\rightarrow d}$ and (d) $S_{d\rightarrow a}$ 
    as a function of probe intensity parameter $\alpha$ and phase mismatch $\Delta{k}_{D}$ 
    for $gz=0.0165$ and $gz=0.067$, respectively.
    The dashed red line corresponds to the threshold of negativity of
    the steering witness. Here, $a\equiv a_j$ corresponding to both the pump modes. All the quantities shown here and in the rest of
    the plots are dimensionless.}
\end{figure}

The asymmetric nature of steering is visible here as pump mode
can steer anti-Stokes for $gz\gtrapprox0.0165$ but anti-Stokes can
steer pump mode only after a long interaction length $gz\gtrapprox0.067$.
The effect of probe on $S_{d\rightarrow a}$ is less than that on
the pump mode steering anti-Stokes $S_{a\rightarrow d}$ as the pump
mode was directly interacting with the probe. For instance, the effect
of the probe mode is observed only in the stimulated case (when $\alpha\neq0$)
and the effect of the phase parameter of the coherent pumping is not
very significant for $S_{d\rightarrow a}$. The phase mismatch between
the probe and hyper-Raman pump modes $\Delta k_{D}$ can be pertinent
in wiping out both kinds of steering observed (see Fig. 2(c) and (d)).
The observed features of steering show different dependencies on the
pump-probe phase mismatch parameter as $S_{d\rightarrow a}$ ($S_{a\rightarrow d}$)
is symmetric (not symmetric) along the ideal phase matching condition.
It is imperative to state here that the steering witness is sufficient, but
not necessary, which cannot conclude that steering could not be observed
in the other modes we failed to observe here. One such example could
be Stokes--anti-Stokes steering reported under strong coherent pump
condition \citep{KT-PRA}, but not observed here.

\subsection{Quantum Entanglement}

{   The simplest way of defining quantum entanglement is that the
quantum state of a composite system cannot be expressed as a tensor
product of the states of the sub-systems. There are several moment-based
criteria \citep{HZ-criteria,HZ-criteria1,HZ-criteria2,Duan-criteria}
to investigate intermodal quantum entanglement in the bosonic systems.
Since each of the criteria we employed in this study is only sufficient
(i.e., these are not necessary criteria) to identify the existence
of entanglement, the failure of a criterion in detecting the presence
of intermodal entanglement does not imply the absence of entanglement.
It is probable that for a set of state parameters, one criterion fails
to detect entanglement but another criterion detects the same. Keeping
this possibility in mind, in what follows, we have used two criteria
for investigating the existence of intermodal entanglement in the
probed-hyper-Raman system of our interest. These criteria are provided
by Hillery and Zubairy \citep{HZ-criteria,HZ-criteria1,HZ-criteria2};
and henceforth, we will refer to these two criteria as HZ-1 and HZ-2
criteria. Specifically, these two criteria can be expressed as }{ }

{  
\begin{equation}
\begin{array}{lcl}
E_{ij} & = & \left\langle N_{i}N_{j}\right\rangle -\left|\left\langle ij^{\dagger}\right\rangle \right|^{2}<0,\end{array}\label{HZ-1=000020criteria}
\end{equation}
}{ }

{   and
\begin{equation}
\begin{array}{lcl}
E_{ij}^{\prime} & = & \left\langle N_{i}\right\rangle \left\langle N_{j}\right\rangle -\left|\left\langle ij\right\rangle \right|^{2}<0,\end{array}\label{HZ-2=000020criteria}
\end{equation}
where $i$ and $j$ are any two arbitrary operators and $N_{i\,}(N_{j})=i^{\dagger}i\,(j^{\dagger}j)$
are the number operators for $i^{th}\,(j^{th})$ mode and $i\neq j$. The expressions of witnesses for all possible intermodal entanglement are given in Appendix B.
Here, it would be apt to note that if two modes are found to show
steering they must be entangled but the converse is not true for mixed
state. As the presence of steering implies the presence of entanglement,
the existence of quantum steering in pump-anti-Stokes mode is the signature
of quantum entanglement in these two modes. However, unlike steering,
this compound mode remains always entangled. This shows that switching
between steering state and entangled state can be obtained by controlling
different interaction parameters discussed previously. We also observed
that the depth of the entanglement witness in this case can be regulated
by the presence of the probe as it becomes less negative. Although
this does not lead to the inseparability of pump and anti-Stokes mode,
it is expected to reduce the amount of entanglement when quantified
using nonclassicality depth \citep{Lee-depth} as performed in (\citep{Exp}
and references therein). In contrast, Stokes-phonon and Stokes-anti-Stokes
entanglement witnessed by HZ-2 and HZ-1 criterion remains unaltered
due to the presence of the probe. }{ }

\begin{figure}[t]
    \centering
    \includegraphics[scale=0.33]{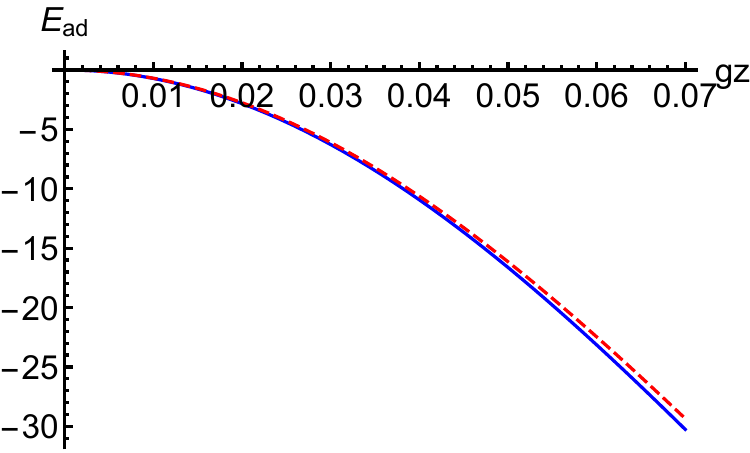}
    \includegraphics[scale=0.33]{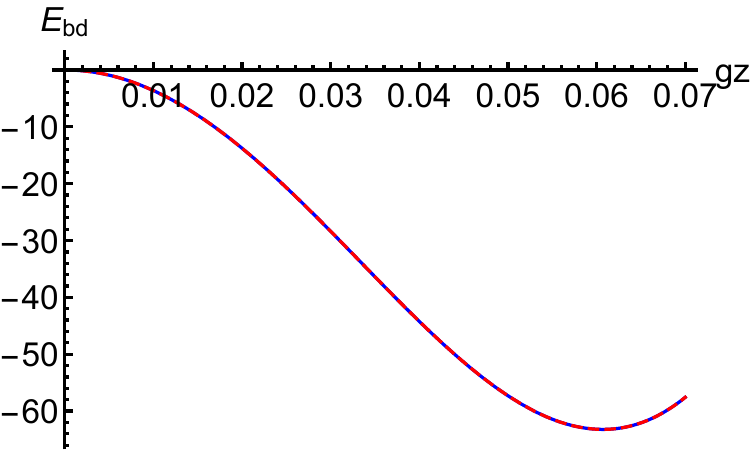}
    \includegraphics[scale=0.33]{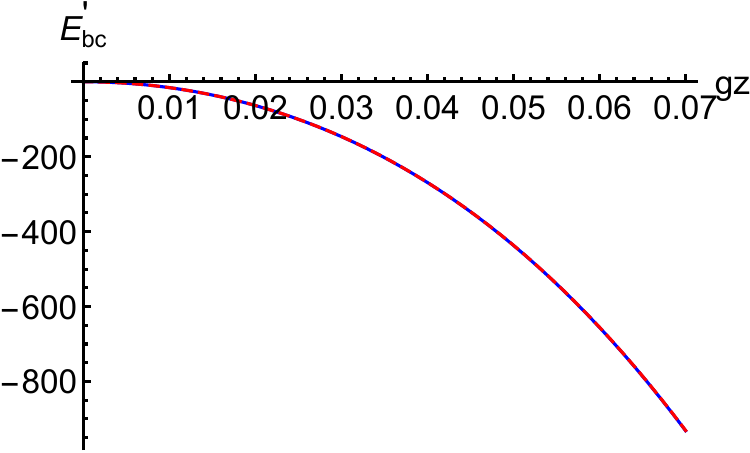}
    \centerline{ \hspace{2mm} (a) \hspace{.3\hsize} (b)  \hspace{.3\hsize} (c) }\\
    
    \caption{\protect\label{fig:Ent} Variation of entanglement
    characterized by (a) HZ-1 criterion in pump-anti-Stokes, (b) HZ-1
    criterion in Stokes-anti-Stokes, and (c) HZ-2 criterion in Stokes-phonon
    modes for $\Gamma=0$ and $\Gamma=1.5g$, shown in smooth (blue) and
    dashed (red) lines, respectively. Stokes-anti-Stokes entanglement
    in (b) is only observed for the phase angle $\phi_{1}=\frac{\pi}{2}$
    for the pump. The rest of the parameters are the same as in the previous
    figure.}
\end{figure}

\subsection{Antibunching}
Unlike standard light sources, which can emit photons in bunched
formations, antibunched photons are detected individually, with temporal
intervals between detections. This nonclassical behavior is a key
notion in quantum optics and information science, and it serves as
the foundation for a wide range of advanced quantum technologies.
Bunching or antibunching of light can be understood with the second-order
correlation function. The second order correlational function $g^{(2)}(0)$
is defined as
\begin{equation}
\begin{array}{lcl}
g_{i}^{(2)}(0) & = & \frac{\left\langle i^{\dagger}(z)i^{\dagger}(z)i(z)i(z)\right\rangle }{\left\langle i^{\dagger}(z)i(z)\right\rangle \left\langle i^{\dagger}(z)i(z)\right\rangle }\end{array},\label{eq:secondordercorrelation}
\end{equation}
where $i$ denotes the $i^{th}$mode $\left(i:i\in\left\{ a_{p},a_{1},a_{2},\,b,\,c,\,d\right\} \right)$.
After rearranging Eq.(\ref{eq:secondordercorrelation}), we obtain
\begin{equation}
\begin{array}{lcl}
g_{i}^{(2)}(0) & = & 1+\frac{\left\langle \left(\Delta N_{i}\right)^{2}\right\rangle -\left\langle N_{i}\right\rangle }{\left\langle N_{i}\right\rangle ^{2}}\\
 & = & 1+\frac{D_{i}}{\left\langle N_{i}\right\rangle ^{2}},
\end{array}\label{eq:secondordercorrelation2}
\end{equation}
where $D_{i}$ is the deference between the variance $\left\langle \left(\Delta N_{i}\right)^{2}\right\rangle $
and $\left\langle N_{i}\right\rangle $ with $\Delta N_{i}=\left(N_{i}-\left\langle N_{i}\right\rangle \right)$
and $N_{i}=i^{\dagger}i$ the number operator for the $i^{th}$mode.
As $g_{i}^{(2)}(0)<1$ correspond to $D_{i}<0$ in Eq. (\ref{eq:secondordercorrelation2})
for the $i^{th}$ mode reveals photon antibunching. The expressions of witnesses for all possible photon antibunching are given in Appendix B, where we can observe that antibuching could not be observed in Stokes and anti-Stokes modes.

\begin{figure}[t]
    \centering
    \includegraphics[scale=0.6]{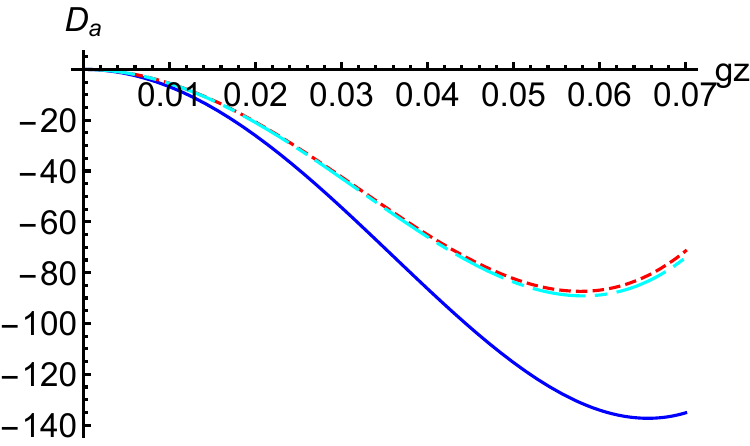}
    \caption{\protect\label{fig:Ant} (Color online) Photon antibunching is observed in pump mode for $\Gamma=0$, $\alpha\neq0$ or $\Gamma\neq0$, $\alpha=0$, $\Gamma=1.5g,\alpha=9$, and $\Gamma=1.5g,\alpha=9\exp(i\pi)$ shown in smooth (blue), dashed (red), and dot-dashed (cyan) lines, respectively. Antibunching is only observed for the phase angle $\phi_{1}=\frac{\pi}{2}$ for the pump and not $\phi_{1}=0$. The rest of the parameters are the same as in the previous figure.}
\end{figure}

{   The pump mode photon antibunching observed in the hyper-Raman
process undergoing in the system waveguide depends on the probe interaction
parameters (cf. Fig. \ref{fig:Ant}). The depth of the nonclassicality
witness decreases due to the presence of the probe initially in the
coherent state. The probe mode in the absence of the coherent pumping
does not impact the photon antibunching observed in the pump modes.
The depth of the witness also depends on the phase parameter of the
coherent state. Therefore, all these parameters could impact the nonclassicality
depth of the pump mode photon antibunching (\citep{Exp} and references
therein).}{ }

\section{{   Conclusion \label{sec:Conclusion}}}

{   Raman and hyper-Raman processes are known as good candidates
for the generation of quantum correlated states. Here, we show that this
process of generation of correlated states can be controlled by an
external probe interacting nonlinearly with both the non-degenerate
pump modes of the hyper-Raman process. We have shown the existence
of quantum steering between pump and anti-Stokes modes, which can
be influenced significantly by controlling the interaction between
the system and the probe. This may be relevant in the implementation
of one-way device independent quantum technology based on a quantum
steering state. For instance, if an eavesdropper can induce switching
between quantum steering state to entangled state by the probe, it
can compromise the device independent security. }{ }

{   In addition to quantum steering and the possibilities of the
generation of entangled and antibunched states in the above described
probed-hyper-Raman system is also studied with a specific focus on
the impact of the external probe (specifically, the impact of the
probe interaction parameters and the parameters that define the initial
state of the pump (e.g., phase parameter of the coherent state if
the system is considered to be coherently pumped). No signature of
antibunching is observed in Stokes and anti-Stokes modes, and it was
found that in these cases, the witness of antibunching is free from
the probe interaction term. However, photon antibunching is observed
in the pump mode, and the same is found to depend on the probe interaction
parameters. Interestingly, it is observed that the impact of the probe
on the photon antibunching characteristics of the pump mode vanishes
in the absence of coherent pumping. The depth of the nonclassicality
(antibunching) witness for the antibunching observed in the pump mode
of the coherently pumped probed-hyper-Raman system is also found to
depend on the phase parameter of the coherent state. }{ }

{   We expect the present study will motivate more studies dedicated
to probing the nonclassicality generation from a quantum system. For
instance, an asymmetric nonlinear optical coupler operating under
non-degenerate hyper-Raman process will be studied in the future with
contra-propagating probe mode \citep{kishore2014contra} to exploit
the efficient coupling between the waveguides.}{ }

\bmhead{Acknowledgements}
KT acknowledges the financial
support from the Operational Programme Research, Development and Education
- European Regional Development Fund project no. OP JAC CZ.02.01.01/00/22\_008/0004596
of the Ministry of Education, Youth and Sports of the Czech Republic.

\section*{Declarations}
\begin{itemize}
\item There is no data given in this paper as such, though the data corresponding to the parameters in this article is available in the public domain.
\item All the authors have completely conceptualised the problem and written \& edited the draft.
\end{itemize}
\noindent

\bibliography{sn-bibliography}


\begin{thebibliography}{37}
\ifx \bisbn   \undefined \def \bisbn  #1{ISBN #1}\fi
\ifx \binits  \undefined \def \binits#1{#1}\fi
\ifx \bauthor  \undefined \def \bauthor#1{#1}\fi
\ifx \batitle  \undefined \def \batitle#1{#1}\fi
\ifx \bjtitle  \undefined \def \bjtitle#1{#1}\fi
\ifx \bvolume  \undefined \def \bvolume#1{\textbf{#1}}\fi
\ifx \byear  \undefined \def \byear#1{#1}\fi
\ifx \bissue  \undefined \def \bissue#1{#1}\fi
\ifx \bfpage  \undefined \def \bfpage#1{#1}\fi
\ifx \blpage  \undefined \def \blpage #1{#1}\fi
\ifx \burl  \undefined \def \burl#1{\textsf{#1}}\fi
\ifx \doiurl  \undefined \def \doiurl#1{\url{https://doi.org/#1}}\fi
\ifx \betal  \undefined \def \betal{\textit{et al.}}\fi
\ifx \binstitute  \undefined \def \binstitute#1{#1}\fi
\ifx \binstitutionaled  \undefined \def \binstitutionaled#1{#1}\fi
\ifx \bctitle  \undefined \def \bctitle#1{#1}\fi
\ifx \beditor  \undefined \def \beditor#1{#1}\fi
\ifx \bpublisher  \undefined \def \bpublisher#1{#1}\fi
\ifx \bbtitle  \undefined \def \bbtitle#1{#1}\fi
\ifx \bedition  \undefined \def \bedition#1{#1}\fi
\ifx \bseriesno  \undefined \def \bseriesno#1{#1}\fi
\ifx \blocation  \undefined \def \blocation#1{#1}\fi
\ifx \bsertitle  \undefined \def \bsertitle#1{#1}\fi
\ifx \bsnm \undefined \def \bsnm#1{#1}\fi
\ifx \bsuffix \undefined \def \bsuffix#1{#1}\fi
\ifx \bparticle \undefined \def \bparticle#1{#1}\fi
\ifx \barticle \undefined \def \barticle#1{#1}\fi
\bibcommenthead
\ifx \bconfdate \undefined \def \bconfdate #1{#1}\fi
\ifx \botherref \undefined \def \botherref #1{#1}\fi
\ifx \url \undefined \def \url#1{\textsf{#1}}\fi
\ifx \bchapter \undefined \def \bchapter#1{#1}\fi
\ifx \bbook \undefined \def \bbook#1{#1}\fi
\ifx \bcomment \undefined \def \bcomment#1{#1}\fi
\ifx \oauthor \undefined \def \oauthor#1{#1}\fi
\ifx \citeauthoryear \undefined \def \citeauthoryear#1{#1}\fi
\ifx \endbibitem  \undefined \def \endbibitem {}\fi
\ifx \bconflocation  \undefined \def \bconflocation#1{#1}\fi
\ifx \arxivurl  \undefined \def \arxivurl#1{\textsf{#1}}\fi
\csname PreBibitemsHook\endcsname

\bibitem[\protect\citeauthoryear{Dowling and Milburn}{2003}]{QT}
\begin{barticle}
\bauthor{\bsnm{Dowling}, \binits{J.P.}},
\bauthor{\bsnm{Milburn}, \binits{G.J.}}:
\batitle{Quantum technology: the second quantum revolution}.
\bjtitle{Philosophical Transactions of the Royal Society of London. Series A:
  Mathematical, Physical and Engineering Sciences}
\bvolume{361}(\bissue{1809}),
\bfpage{1655}--\blpage{1674}
(\byear{2003})
\end{barticle}
\endbibitem

\bibitem[\protect\citeauthoryear{Schleich et~al.}{2016}]{QT-2}
\begin{barticle}
\bauthor{\bsnm{Schleich}, \binits{W.P.}},
\bauthor{\bsnm{Ranade}, \binits{K.S.}},
\bauthor{\bsnm{Anton}, \binits{C.}},
\bauthor{\bsnm{Arndt}, \binits{M.}},
\bauthor{\bsnm{Aspelmeyer}, \binits{M.}},
\bauthor{\bsnm{Bayer}, \binits{M.}},
\bauthor{\bsnm{Berg}, \binits{G.}},
\bauthor{\bsnm{Calarco}, \binits{T.}},
\bauthor{\bsnm{Fuchs}, \binits{H.}},
\bauthor{\bsnm{Giacobino}, \binits{E.}}, \betal:
\batitle{Quantum technology: from research to application}.
\bjtitle{Applied Physics B}
\bvolume{122},
\bfpage{1}--\blpage{31}
(\byear{2016})
\end{barticle}
\endbibitem

\bibitem[\protect\citeauthoryear{Bennett et~al.}{1993}]{Bennet1993}
\begin{barticle}
\bauthor{\bsnm{Bennett}, \binits{C.H.}},
\bauthor{\bsnm{Brassard}, \binits{G.}},
\bauthor{\bsnm{Cr{\'e}peau}, \binits{C.}},
\bauthor{\bsnm{Jozsa}, \binits{R.}},
\bauthor{\bsnm{Peres}, \binits{A.}},
\bauthor{\bsnm{Wootters}, \binits{W.K.}}:
\batitle{Teleporting an unknown quantum state via dual classical and
  {Einstein-Podolsky-Rosen} channels}.
\bjtitle{Physical Review Letters}
\bvolume{70}(\bissue{13}),
\bfpage{1895}
(\byear{1993})
\end{barticle}
\endbibitem

\bibitem[\protect\citeauthoryear{Bennett and Wiesner}{1992}]{densecoding}
\begin{barticle}
\bauthor{\bsnm{Bennett}, \binits{C.H.}},
\bauthor{\bsnm{Wiesner}, \binits{S.J.}}:
\batitle{Communication via one-and two-particle operators on
  {Einstein-Podolsky-Rosen} states}.
\bjtitle{Physical Review Letters}
\bvolume{69}(\bissue{20}),
\bfpage{2881}
(\byear{1992})
\end{barticle}
\endbibitem

\bibitem[\protect\citeauthoryear{Bennett et~al.}{2001}]{RSP}
\begin{barticle}
\bauthor{\bsnm{Bennett}, \binits{C.H.}},
\bauthor{\bsnm{DiVincenzo}, \binits{D.P.}},
\bauthor{\bsnm{Shor}, \binits{P.W.}},
\bauthor{\bsnm{Smolin}, \binits{J.A.}},
\bauthor{\bsnm{Terhal}, \binits{B.M.}},
\bauthor{\bsnm{Wootters}, \binits{W.K.}}:
\batitle{Remote state preparation}.
\bjtitle{Physical Review Letters}
\bvolume{87}(\bissue{7}),
\bfpage{077902}
(\byear{2001})
\end{barticle}
\endbibitem

\bibitem[\protect\citeauthoryear{Acin et~al.}{2006}]{DI-QKD}
\begin{barticle}
\bauthor{\bsnm{Acin}, \binits{A.}},
\bauthor{\bsnm{Gisin}, \binits{N.}},
\bauthor{\bsnm{Masanes}, \binits{L.}}:
\batitle{From {Bell}'s theorem to secure quantum key distribution}.
\bjtitle{Physical Review Letters}
\bvolume{97}(\bissue{12}),
\bfpage{120405}
(\byear{2006})
\end{barticle}
\endbibitem

\bibitem[\protect\citeauthoryear{Liu et~al.}{2018}]{QRNG}
\begin{barticle}
\bauthor{\bsnm{Liu}, \binits{Y.}},
\bauthor{\bsnm{Zhao}, \binits{Q.}},
\bauthor{\bsnm{Li}, \binits{M.-H.}},
\bauthor{\bsnm{Guan}, \binits{J.-Y.}},
\bauthor{\bsnm{Zhang}, \binits{Y.}},
\bauthor{\bsnm{Bai}, \binits{B.}},
\bauthor{\bsnm{Zhang}, \binits{W.}},
\bauthor{\bsnm{Liu}, \binits{W.-Z.}},
\bauthor{\bsnm{Wu}, \binits{C.}},
\bauthor{\bsnm{Yuan}, \binits{X.}}, \betal:
\batitle{Device-independent quantum random-number generation}.
\bjtitle{Nature}
\bvolume{562}(\bissue{7728}),
\bfpage{548}--\blpage{551}
(\byear{2018})
\end{barticle}
\endbibitem

\bibitem[\protect\citeauthoryear{Sen et~al.}{2011}]{bsen-subshot}
\begin{barticle}
\bauthor{\bsnm{Sen}, \binits{B.}},
\bauthor{\bsnm{Pe{\v{r}}inov{\'a}}, \binits{V.}},
\bauthor{\bsnm{Luk{\v{s}}}, \binits{A.}},
\bauthor{\bsnm{Pe{\v{r}}ina}, \binits{J.}},
\bauthor{\bsnm{K{\v{r}}epelka}, \binits{J.}}:
\batitle{Sub-shot noise photon-number correlation in stimulated and spontaneous
  {Raman} processes}.
\bjtitle{Journal of Physics B: Atomic, Molecular and Optical Physics}
\bvolume{44}(\bissue{10}),
\bfpage{105503}
(\byear{2011})
\end{barticle}
\endbibitem

\bibitem[\protect\citeauthoryear{Sen et~al.}{2007}]{bsen-mandal-hyper}
\begin{barticle}
\bauthor{\bsnm{Sen}, \binits{B.}},
\bauthor{\bsnm{Mandal}, \binits{S.}},
\bauthor{\bsnm{Perina}, \binits{J.}}:
\batitle{Quantum statistical properties of the radiation field in spontaneous
  {Raman} and stimulated {Raman} processes}.
\bjtitle{Journal of Physics B: Atomic, Molecular and Optical Physics}
\bvolume{40}(\bissue{7}),
\bfpage{1417}
(\byear{2007})
\end{barticle}
\endbibitem

\bibitem[\protect\citeauthoryear{Thapliyal and
  Pe{\v{r}}ina}{2019}]{off-res-Raman}
\begin{barticle}
\bauthor{\bsnm{Thapliyal}, \binits{K.}},
\bauthor{\bsnm{Pe{\v{r}}ina}, \binits{J.}}:
\batitle{Nonclassicality in off-resonant {Raman} process}.
\bjtitle{Physics Letters A}
\bvolume{383}(\bissue{17}),
\bfpage{2011}--\blpage{2020}
(\byear{2019})
\end{barticle}
\endbibitem

\bibitem[\protect\citeauthoryear{Thapliyal and
  Pe{\v{r}}ina}{2020}]{off-res-RamanPh}
\begin{barticle}
\bauthor{\bsnm{Thapliyal}, \binits{K.}},
\bauthor{\bsnm{Pe{\v{r}}ina}, \binits{J.}}:
\batitle{Quasidistribution of phases in {Raman} process with weak and strong
  pumps}.
\bjtitle{Physica Scripta}
\bvolume{95}(\bissue{3}),
\bfpage{034001}
(\byear{2020})
\end{barticle}
\endbibitem

\bibitem[\protect\citeauthoryear{Thapliyal and Pe{\v{r}}ina~Jr}{2021}]{KT-PRA}
\begin{barticle}
\bauthor{\bsnm{Thapliyal}, \binits{K.}},
\bauthor{\bsnm{Pe{\v{r}}ina~Jr}, \binits{J.}}:
\batitle{Ideal pairing of the {Stokes} and anti-{Stokes} photons in the {Raman}
  process}.
\bjtitle{Physical Review A}
\bvolume{103}(\bissue{3}),
\bfpage{033708}
(\byear{2021})
\end{barticle}
\endbibitem

\bibitem[\protect\citeauthoryear{Sen et~al.}{2013}]{giri1}
\begin{barticle}
\bauthor{\bsnm{Sen}, \binits{B.}},
\bauthor{\bsnm{Giri}, \binits{S.K.}},
\bauthor{\bsnm{Mandal}, \binits{S.}},
\bauthor{\bsnm{Ooi}, \binits{C.R.}},
\bauthor{\bsnm{Pathak}, \binits{A.}}:
\batitle{Intermodal entanglement in {Raman} processes}.
\bjtitle{Physical Review A}
\bvolume{87}(\bissue{2}),
\bfpage{022325}
(\byear{2013})
\end{barticle}
\endbibitem

\bibitem[\protect\citeauthoryear{Giri et~al.}{2016}]{giri2}
\begin{barticle}
\bauthor{\bsnm{Giri}, \binits{S.K.}},
\bauthor{\bsnm{Sen}, \binits{B.}},
\bauthor{\bsnm{Pathak}, \binits{A.}},
\bauthor{\bsnm{Jana}, \binits{P.C.}}:
\batitle{Higher-order two-mode and multimode entanglement in {Raman}
  processes}.
\bjtitle{Physical Review A}
\bvolume{93}(\bissue{1}),
\bfpage{012340}
(\byear{2016})
\end{barticle}
\endbibitem

\bibitem[\protect\citeauthoryear{Pe\v{r}ina}{1991}]{perina-book}
\begin{bbook}
\bauthor{\bsnm{Pe\v{r}ina}, \binits{J.}}:
\bbtitle{Quantum Statistics of Linear and Nonlinear Optical Phenomena}.
\bpublisher{Kluwer},
\blocation{Dordrecht}
(\byear{1991})
\end{bbook}
\endbibitem

\bibitem[\protect\citeauthoryear{Dochow et~al.}{2013}]{Raman-chip}
\begin{barticle}
\bauthor{\bsnm{Dochow}, \binits{S.}},
\bauthor{\bsnm{Becker}, \binits{M.}},
\bauthor{\bsnm{Spittel}, \binits{R.}},
\bauthor{\bsnm{Beleites}, \binits{C.}},
\bauthor{\bsnm{Stanca}, \binits{S.}},
\bauthor{\bsnm{Latka}, \binits{I.}},
\bauthor{\bsnm{Schuster}, \binits{K.}},
\bauthor{\bsnm{Kobelke}, \binits{J.}},
\bauthor{\bsnm{Unger}, \binits{S.}},
\bauthor{\bsnm{Henkel}, \binits{T.}}, \betal:
\batitle{Raman-on-chip device and detection fibres with fibre {Bragg} grating
  for analysis of solutions and particles}.
\bjtitle{Lab on a Chip}
\bvolume{13}(\bissue{6}),
\bfpage{1109}--\blpage{1113}
(\byear{2013})
\end{barticle}
\endbibitem

\bibitem[\protect\citeauthoryear{Dhakal et~al.}{2016}]{Raman-chip-rev}
\begin{barticle}
\bauthor{\bsnm{Dhakal}, \binits{A.}},
\bauthor{\bsnm{Peyskens}, \binits{F.}},
\bauthor{\bsnm{Clemmen}, \binits{S.}},
\bauthor{\bsnm{Raza}, \binits{A.}},
\bauthor{\bsnm{Wuytens}, \binits{P.}},
\bauthor{\bsnm{Zhao}, \binits{H.}},
\bauthor{\bsnm{Le~Thomas}, \binits{N.}},
\bauthor{\bsnm{Baets}, \binits{R.}}:
\batitle{Single mode waveguide platform for spontaneous and surface-enhanced
  on-chip {Raman} spectroscopy}.
\bjtitle{Interface focus}
\bvolume{6}(\bissue{4}),
\bfpage{20160015}
(\byear{2016})
\end{barticle}
\endbibitem

\bibitem[\protect\citeauthoryear{Kasperczyk et~al.}{2016}]{Raman-Q}
\begin{barticle}
\bauthor{\bsnm{Kasperczyk}, \binits{M.}},
\bauthor{\bsnm{Aguiar~J{\'u}nior}, \binits{F.S.}},
\bauthor{\bsnm{Rabelo}, \binits{C.}},
\bauthor{\bsnm{Saraiva}, \binits{A.}},
\bauthor{\bsnm{Santos}, \binits{M.F.}},
\bauthor{\bsnm{Novotny}, \binits{L.}},
\bauthor{\bsnm{Jorio}, \binits{A.}}:
\batitle{Temporal quantum correlations in inelastic light scattering from
  water}.
\bjtitle{Physical Review Letters}
\bvolume{117}(\bissue{24}),
\bfpage{243603}
(\byear{2016})
\end{barticle}
\endbibitem

\bibitem[\protect\citeauthoryear{Duan et~al.}{2001}]{Raman-Q1}
\begin{barticle}
\bauthor{\bsnm{Duan}, \binits{L.-M.}},
\bauthor{\bsnm{Lukin}, \binits{M.D.}},
\bauthor{\bsnm{Cirac}, \binits{J.I.}},
\bauthor{\bsnm{Zoller}, \binits{P.}}:
\batitle{Long-distance quantum communication with atomic ensembles and linear
  optics}.
\bjtitle{Nature}
\bvolume{414}(\bissue{6862}),
\bfpage{413}--\blpage{418}
(\byear{2001})
\end{barticle}
\endbibitem

\bibitem[\protect\citeauthoryear{Das et~al.}{2021}]{das}
\begin{barticle}
\bauthor{\bsnm{Das}, \binits{M.}},
\bauthor{\bsnm{Thapliyal}, \binits{K.}},
\bauthor{\bsnm{Sen}, \binits{B.}},
\bauthor{\bsnm{Pe{\v{r}}ina}, \binits{J.}},
\bauthor{\bsnm{Pathak}, \binits{A.}}:
\batitle{Interplay between quantum {Zeno} and anti-{Zeno} effects in a
  nondegenerate hyper-{Raman} nonlinear optical coupler}.
\bjtitle{Physical Review A}
\bvolume{103}(\bissue{1}),
\bfpage{013713}
(\byear{2021})
\end{barticle}
\endbibitem

\bibitem[\protect\citeauthoryear{Das et~al.}{2023}]{das2}
\begin{barticle}
\bauthor{\bsnm{Das}, \binits{M.}},
\bauthor{\bsnm{Sen}, \binits{B.}},
\bauthor{\bsnm{Thapliyal}, \binits{K.}},
\bauthor{\bsnm{Pathak}, \binits{A.}}:
\batitle{Quantum {Zeno} and anti-{Zeno} effects in the dynamics of
  non-degenerate hyper-{Raman} processes coupled to two linear waveguides}.
\bjtitle{Annalen der Physik}
\bvolume{535}(\bissue{10}),
\bfpage{2300196}
(\byear{2023})
\end{barticle}
\endbibitem

\bibitem[\protect\citeauthoryear{Thapliyal et~al.}{2016}]{NZeno}
\begin{barticle}
\bauthor{\bsnm{Thapliyal}, \binits{K.}},
\bauthor{\bsnm{Pathak}, \binits{A.}},
\bauthor{\bsnm{Pe{\v{r}}ina}, \binits{J.}}:
\batitle{Linear and nonlinear quantum {Zeno} and anti-{Zeno} effects in a
  nonlinear optical coupler}.
\bjtitle{Physical Review A}
\bvolume{93}(\bissue{2}),
\bfpage{022107}
(\byear{2016})
\end{barticle}
\endbibitem

\bibitem[\protect\citeauthoryear{Thun et~al.}{2002}]{thun2002zeno-raman}
\begin{barticle}
\bauthor{\bsnm{Thun}, \binits{K.}},
\bauthor{\bsnm{Pe{\v{r}}ina}, \binits{J.}},
\bauthor{\bsnm{K{\v{r}}epelka}, \binits{J.}}:
\batitle{Quantum {Zeno} effect in {Raman} scattering}.
\bjtitle{Physics Letters A}
\bvolume{299}(\bissue{1}),
\bfpage{19}--\blpage{30}
(\byear{2002})
\end{barticle}
\endbibitem

\bibitem[\protect\citeauthoryear{Thapliyal
  et~al.}{2014}]{kishore2014co-coupler}
\begin{barticle}
\bauthor{\bsnm{Thapliyal}, \binits{K.}},
\bauthor{\bsnm{Pathak}, \binits{A.}},
\bauthor{\bsnm{Sen}, \binits{B.}},
\bauthor{\bsnm{Pe{\v{r}}ina}, \binits{J.}}:
\batitle{Higher-order nonclassicalities in a codirectional nonlinear optical
  coupler: Quantum entanglement, squeezing, and antibunching}.
\bjtitle{Physical Review A}
\bvolume{90}(\bissue{1}),
\bfpage{013808}
(\byear{2014})
\end{barticle}
\endbibitem

\bibitem[\protect\citeauthoryear{Gisin}{1991}]{gisin}
\begin{barticle}
\bauthor{\bsnm{Gisin}, \binits{N.}}:
\batitle{Bell's inequality holds for all non-product states}.
\bjtitle{Physics Letters A}
\bvolume{154}(\bissue{5-6}),
\bfpage{201}--\blpage{202}
(\byear{1991})
\end{barticle}
\endbibitem

\bibitem[\protect\citeauthoryear{Popescu and Rohrlich}{1992}]{popescu}
\begin{barticle}
\bauthor{\bsnm{Popescu}, \binits{S.}},
\bauthor{\bsnm{Rohrlich}, \binits{D.}}:
\batitle{Generic quantum nonlocality}.
\bjtitle{Physics Letters A}
\bvolume{166}(\bissue{5-6}),
\bfpage{293}--\blpage{297}
(\byear{1992})
\end{barticle}
\endbibitem

\bibitem[\protect\citeauthoryear{Branciard et~al.}{2012}]{1way-QKD}
\begin{barticle}
\bauthor{\bsnm{Branciard}, \binits{C.}},
\bauthor{\bsnm{Cavalcanti}, \binits{E.G.}},
\bauthor{\bsnm{Walborn}, \binits{S.P.}},
\bauthor{\bsnm{Scarani}, \binits{V.}},
\bauthor{\bsnm{Wiseman}, \binits{H.M.}}:
\batitle{One-sided device-independent quantum key distribution: Security,
  feasibility, and the connection with steering}.
\bjtitle{Physical Review A}
\bvolume{85}(\bissue{1}),
\bfpage{010301}
(\byear{2012})
\end{barticle}
\endbibitem

\bibitem[\protect\citeauthoryear{He et~al.}{2012}]{steeringcriteria-1}
\begin{barticle}
\bauthor{\bsnm{He}, \binits{Q.}},
\bauthor{\bsnm{Drummond}, \binits{P.}},
\bauthor{\bsnm{Olsen}, \binits{M.}},
\bauthor{\bsnm{Reid}, \binits{M.}}:
\batitle{{Einstein-Podolsky-Rosen} entanglement and steering in two-well
  {Bose-Einstein}-condensate ground states}.
\bjtitle{Physical Review A}
\bvolume{86}(\bissue{2}),
\bfpage{023626}
(\byear{2012})
\end{barticle}
\endbibitem

\bibitem[\protect\citeauthoryear{Naikoo et~al.}{2019}]{steeringcriteria-2}
\begin{barticle}
\bauthor{\bsnm{Naikoo}, \binits{J.}},
\bauthor{\bsnm{Thapliyal}, \binits{K.}},
\bauthor{\bsnm{Banerjee}, \binits{S.}},
\bauthor{\bsnm{Pathak}, \binits{A.}}:
\batitle{Quantum {Zeno} effect and nonclassicality in a {PT}-symmetric system
  of coupled cavities}.
\bjtitle{Physical Review A}
\bvolume{99}(\bissue{2}),
\bfpage{023820}
(\byear{2019})
\end{barticle}
\endbibitem

\bibitem[\protect\citeauthoryear{Hillery and Zubairy}{2006a}]{HZ-criteria}
\begin{barticle}
\bauthor{\bsnm{Hillery}, \binits{M.}},
\bauthor{\bsnm{Zubairy}, \binits{M.S.}}:
\batitle{Entanglement conditions for two-mode states}.
\bjtitle{Physical Review Letters}
\bvolume{96}(\bissue{5}),
\bfpage{050503}
(\byear{2006})
\end{barticle}
\endbibitem

\bibitem[\protect\citeauthoryear{Hillery and Zubairy}{2006b}]{HZ-criteria1}
\begin{barticle}
\bauthor{\bsnm{Hillery}, \binits{M.}},
\bauthor{\bsnm{Zubairy}, \binits{M.S.}}:
\batitle{Entanglement conditions for two-mode states: applications}.
\bjtitle{Physical Review A}
\bvolume{74}(\bissue{3}),
\bfpage{032333}
(\byear{2006})
\end{barticle}
\endbibitem

\bibitem[\protect\citeauthoryear{Hillery et~al.}{2010}]{HZ-criteria2}
\begin{barticle}
\bauthor{\bsnm{Hillery}, \binits{M.}},
\bauthor{\bsnm{Dung}, \binits{H.T.}},
\bauthor{\bsnm{Zheng}, \binits{H.}}:
\batitle{Conditions for entanglement in multipartite systems}.
\bjtitle{Physical Review A}
\bvolume{81}(\bissue{6}),
\bfpage{062322}
(\byear{2010})
\end{barticle}
\endbibitem

\bibitem[\protect\citeauthoryear{Duan et~al.}{2000}]{Duan-criteria}
\begin{barticle}
\bauthor{\bsnm{Duan}, \binits{L.-M.}},
\bauthor{\bsnm{Giedke}, \binits{G.}},
\bauthor{\bsnm{Cirac}, \binits{J.I.}},
\bauthor{\bsnm{Zoller}, \binits{P.}}:
\batitle{Inseparability criterion for continuous variable systems}.
\bjtitle{Physical Review Letters}
\bvolume{84}(\bissue{12}),
\bfpage{2722}
(\byear{2000})
\end{barticle}
\endbibitem

\bibitem[\protect\citeauthoryear{Lee}{1991}]{Lee-depth}
\begin{barticle}
\bauthor{\bsnm{Lee}, \binits{C.T.}}:
\batitle{Measure of the nonclassicality of nonclassical states}.
\bjtitle{Physical Review A}
\bvolume{44}(\bissue{5}),
\bfpage{2775}
(\byear{1991})
\end{barticle}
\endbibitem

\bibitem[\protect\citeauthoryear{Thapliyal et~al.}{2024}]{Exp}
\begin{barticle}
\bauthor{\bsnm{Thapliyal}, \binits{K.}},
\bauthor{\bsnm{Pe{\v{r}}ina~Jr}, \binits{J.}},
\bauthor{\bsnm{Haderka}, \binits{O.}},
\bauthor{\bsnm{Mich{\'a}lek}, \binits{V.}},
\bauthor{\bsnm{Machulka}, \binits{R.}}:
\batitle{Experimental characterization of multimode photon-subtracted twin
  beams}.
\bjtitle{Physical Review Research}
\bvolume{6}(\bissue{1}),
\bfpage{013065}
(\byear{2024})
\end{barticle}
\endbibitem

\bibitem[\protect\citeauthoryear{Thapliyal et~al.}{2014}]{kishore2014contra}
\begin{barticle}
\bauthor{\bsnm{Thapliyal}, \binits{K.}},
\bauthor{\bsnm{Pathak}, \binits{A.}},
\bauthor{\bsnm{Sen}, \binits{B.}},
\bauthor{\bsnm{Perina}, \binits{J.}}:
\batitle{Nonclassical properties of a contradirectional nonlinear optical
  coupler}.
\bjtitle{Physics Letters A}
\bvolume{378}(\bissue{46}),
\bfpage{3431}--\blpage{3440}
(\byear{2014})
\end{barticle}
\endbibitem

\bibitem[\protect\citeauthoryear{Pe{\v{r}}inov{\'a} and
  Pe{\v{r}}ina}{1978}]{val-ST}
\begin{barticle}
\bauthor{\bsnm{Pe{\v{r}}inov{\'a}}, \binits{V.}},
\bauthor{\bsnm{Pe{\v{r}}ina}, \binits{J.}}:
\batitle{Generalized {Fokker-Planck} equation approach to optical parametric
  processes {I. Equations} of motion and their solutions}.
\bjtitle{Czechoslovak Journal of Physics B}
\bvolume{28}(\bissue{11}),
\bfpage{1183}--\blpage{1195}
(\byear{1978})
\end{barticle}
\endbibitem

\end{thebibliography}

\appendix

\section*{{   Appendix A: Description of Perturbative Solution\protect\label{sec:app-A}}}

{   The set of coupled differential equations (\ref{eq:equationofmotion})
involving various nonlinear bosonic operators and thus are not exactly
solvable in the closed analytical form under weak pump condition.
Hence, we used Sen-Mandal perturbative technique which is more general
than the well-known short-time approximation (see \cite{off-res-Raman,kishore2014co-coupler,kishore2014contra} for more discussion). The details of the procedure
followed to obtain the solution is also described in our previous work
\citep{das}. However, to make the article self sufficient, here we
briefly discuss the technique used to obtain the solution. We assume
that the spatial evolution of each mode in the Sen-Mandal perturbative
technique is expressed as the product of all powers of the initial
values of the operators $A(z)=\sum_{i,j,k,l,m,n,o,p,q,r,s,t}\mathcal{F}_{y}a_{p}^{\dagger i}(0)a_{p}^{j}(0)a_{1}^{\dagger k}(0)a_{1}^{l}(0)a_{2}^{\dagger m}(0)a_{2}^{n}(0)$
$\times b^{\dagger o}(0)b^{p}(0)c^{\dagger q}(0)c^{r}(0)d^{\dagger s}(0)d^{t}(0)$.
Although this would generate an infinite series of combination of
operators but we will restrict $\mathcal{F}_{y}$ with not higher
than quadratic terms in $\Lambda t\,\forall\,\Lambda\in\{g,\chi,\Gamma\}$.
For instance, the assumed solution for all the field modes are obtained
as
\begin{equation}
\begin{array}{lcl}
a_{p}(z) & = & f_{1}a_{p}(0)+f_{2}a_{1}(0)a_{2}(0)+f_{3}a_{1}(0)a_{1}^{\dagger}(0)b(0)c(0)+f_{4}a_{2}^{\dagger}(0)a_{2}(0)b(0)c(0)\\
 & + & f_{5}a_{1}(0)a_{1}^{\dagger}(0)c^{\dagger}(0)d(0)+f_{6}a_{2}^{\dagger}(0)a_{2}(0)c^{\dagger}(0)d(0)+f_{7}a_{1}(0)a_{1}^{\dagger}(0)a_{p}(0)\\
 & + & f_{8}a_{2}^{\dagger}(0)a_{2}(0)a_{p}(0)\\
a_{1}(z) & = & g_{1}a_{1}(0)+g_{2}a_{2}^{\dagger}(0)b(0)c(0)+g_{3}a_{2}^{\dagger}(0)c^{\dagger}(0)d(0)+g_{4}a_{p}(0)a_{2}^{\dagger}(0)\\
 &+ & g_{5}a_{1}(0)b(0)c^{2}(0)d^{^{\dagger}}(0)+g_{6}a_{1}(0)b^{\dagger}(0)c^{\dagger2}(0)d(0)+g_{7}a_{p}^{\dagger}(0)a_{1}(0)b(0)c(0)\\
 & + & g_{8}a_{p}(0)a_{1}(0)b^{\dagger}(0)c^{\dagger}(0)+g_{9}a_{p}^{\dagger}(0)a_{1}(0)c^{\dagger}(0)d(0)+g_{_{10}}a_{p}(0)a_{1}(0)c(0)d^{\dagger}(0)\\
 & + & g_{11}a_{1}(0)b^{\dagger}(0)b(0)c^{\dagger}(0)c(0)+g_{12}a_{1}(0)a_{2}^{\dagger}(0)a_{2}(0)b(0)b^{\dagger}(0)\\
 & + & g_{13}a_{1}(0)a_{2}^{\dagger}(0)a_{2}(0)c^{\dagger}(0)c(0)+g_{14}a_{1}(0)a_{2}^{\dagger}(0)a_{2}(0)d^{\dagger}(0)d(0)\\
 & + & g_{15}a_{1}(0)c(0)c^{\dagger}(0)d^{\dagger}(0)d(0)+g_{16}a_{p}^{\dagger}(0)a_{p}(0)a_{1}(0)\\
 & + & g_{17}a_{1}(0)a_{2}^{\dagger}(0)a_{2}(0)c^{\dagger}(0)c(0) + g_{18}a_{1}(0)a_{2}^{\dagger}(0)a_{2}(0)+g_{19}a_{1}^{\dagger}(0)a_{2}^{\dagger2}b(0)d(0)\\
a_{2}(z) & = & h_{1}a_{2}(0)+h_{2}a_{1}^{\dagger}(0)b(0)c(0)+h_{3}a_{1}^{\dagger}(0)c^{\dagger}(0)d(0)+h_{4}a_{p}(0)a_{1}^{\dagger}(0)\\
 & + & h_{5}a_{2}b(0)c^{2}(0)d^{^{\dagger}}(0)+h_{6}a_{2}(0)b^{\dagger}(0)c^{\dagger2}(0)d(0)+h_{7}a_{p}^{\dagger}(0)a_{2}(0)b(0)c(0)\\
 & + & h_{8}a_{p}(0)a_{2}(0)b^{\dagger}(0)c^{\dagger}(0)+ h_{9}a_{p}^{\dagger}(0)a_{2}(0)c^{\dagger}(0)d(0)+h_{_{10}}a_{p}(0)a_{2}(0)c(0)d^{\dagger}(0)\\
 & + & h_{11}a_{2}(0)b^{\dagger}(0)b(0)c^{\dagger}(0)c(0)+h_{12}a_{1}^{\dagger}(0)a_{1}(0)a_{2}(0)b(0)b^{\dagger}(0)\\
 & + & h_{13}a_{1}^{\dagger}(0)a_{1}(0)a_{2}(0)c^{\dagger}(0)c(0)+h_{14}a_{2}(0)a_{1}^{\dagger}(0)a_{1}(0)d^{\dagger}(0)d(0)\\
 & + & h_{15}a_{2}(0)c(0)c^{\dagger}(0)d^{\dagger}(0)d(0)+h_{16}a_{p}^{\dagger}(0)a_{p}(0)a_{2}(0)\\
 & + & h_{17}a_{1}^{\dagger}(0)a_{1}(0)a_{2}(0)c^{\dagger}(0)c(0)+ h_{18}a_{1}^{\dagger}(0)a_{1}(0)a_{2}(0)+h_{19}a_{1}^{\dagger2}(0)a_{2}^{\dagger}(0)b(0)d(0)\\
b(z) & = & j_{1}b(0)+j_{2}a_{1}(0)a_{2}(0)c^{\dagger}(0)+j_{3}a_{1}^{2}(0)a_{2}^{2}(0)d^{\dagger}(0)+j_{4}a_{1}(0)a_{1}^{\dagger}(0)c^{\dagger2}(0)d(0)\\
 & + & j_{5}a_{2}^{\dagger}(0)a_{2}(0)c^{\dagger2}(0)d(0)+j_{6}a_{p}(0)a_{1}(0)a_{1}^{\dagger}(0)c^{\dagger}(0)+j_{7}a_{p}(0)a_{2}^{\dagger}(0)a_{2}(0)c^{\dagger}(0)\\
 & + & j_{8}a_{1}^{\dagger}(0)a_{1}(0)a_{2}^{\dagger}(0)a_{2}(0)b(0)+j_{9}a_{1}(0)a_{1}^{\dagger}(0)b(0)c^{\dagger}(0)c(0)\\
 & + & j_{10}a_{2}^{\dagger}(0)a_{2}(0)b(0)c^{\dagger}(0)c(0)\\
c(z) & = & k_{1}c(0)+k_{2}a_{1}(0)a_{2}(0)b^{\dagger}(0)+k_{3}a_{^{1}}^{\dagger}(0)a_{2}^{\dagger}(0)d(0)+k_{4}a_{1}(0)a_{1}^{\dagger}(0)a_{p}(0)b^{\dagger}(0)\\
 & + & k_{5}a_{2}^{\dagger}(0)a_{2}(0)a_{p}(0)b^{\dagger}(0)+k_{6}a_{1}^{\dagger}(0)a_{1}(0)a_{p}^{\dagger}(0)d(0)+k_{7}a_{2}(0)a_{2}^{\dagger}(0)a_{p}^{\dagger}(0)d(0)\\
 & + & k_{8}a_{1}^{\dagger}(0)a_{1}(0)a_{2}^{\dagger}(0)a_{2}(0)c(0)+k_{9}a_{1}(0)a_{1}^{\dagger}(0)b^{\dagger}(0)b(0)c(0)\\
 & + & k_{10}a_{2}^{\dagger}(0)a_{2}(0)b^{\dagger}(0)b(0)c(0)+ k_{11}a_{1}^{\dagger}(0)a_{1}(0)c(0)d^{\dagger}(0)d(0)\\
 & + & k_{12}a_{2}(0)a_{2}^{\dagger}(0)c(0)d^{\dagger}(0)d(0)+k_{13}a_{1}^{\dagger}(0)a_{1}(0)a_{2}^{\dagger}(0)a_{2}(0)c(0)\\
 & + & k_{14}a_{1}^{\dagger}(0)a_{1}(0)b^{\dagger}(0)c^{\dagger}(0)d(0)+k_{15}a_{2}^{\dagger}(0)a_{2}(0)b^{\dagger}(0)c^{\dagger}(0)d(0)\\
 & + & k_{16}b^{\dagger}(0)c^{\dagger}(0)d(0)\\
d(z) & = & l_{1}d(0)+l_{2}a_{1}(0)a_{2}(0)c(0)+l_{3}a_{1}(0)a_{1}^{\dagger}(0)b(0)c^{2}(0)+l_{4}a_{2}^{\dagger}(0)a_{2}(0)b(0)c^{2}(0)\\
 & + & l_{5}a_{1}^{2}(0)a_{2}^{2}(0)b^{\dagger}(0)+l_{6}a_{1}(0)a_{1}^{\dagger}(0)a_{p}(0)c(0)+l_{7}a_{2}^{\dagger}(0)a_{2}(0)a_{p}(0)c(0)\\
 & + & l_{8}a_{1}(0)a_{1}^{\dagger}(0)c^{\dagger}(0)c(0)d(0)+l_{9}a_{2}^{\dagger}(0)a_{2}(0)c^{\dagger}(0)c(0)d(0)\\
 & + & l_{10}a_{1}(0)a_{1}^{\dagger}(0)a_{2}(0)a_{2}^{\dagger}(0)d(0)
\end{array}\label{eq:assumedsol}
\end{equation}
}{ }

{The coefficients $f_{n}$ we use in equation
(\ref{eq:assumedsol}), arise from the the spatial evolution
of the probe mode $a_{p}(z)$, defined as 
\begin{equation}
\begin{array}{lcl}
a_{p}(z) & = & e^{iGz}a_{p}(0)e^{-iGz}\\
 & = & a_{p}(0)+iz\left[G,a_{p}(0)\right]+\frac{\left(iz\right)^{2}}{2!}\left[G,\left[G,a_{p}(0)\right]\right]+\ldots,
\end{array}
\end{equation}
where all the nested commutators are obtained as 
\begin{equation}
\begin{array}{lcl}
\left[G,a_{p}\left(0\right)\right] & = & -\left(k_{p}a_{p}+\varGamma a_{1}a_{2}\right)\end{array}
\end{equation}
}{ }

{  
\begin{equation}
\begin{array}{lcl}
\left[G,\left[G,a_{p}\left(0\right)\right]\right] & = & k_{p}^{2}a_{p}+\varGamma\left(k_{p}+k_{a_{1}}+k_{a_{2}}\right)a_{1}a_{2}+g\varGamma a_{1}a_{1}^{\dagger}bc\\
 & + & g\varGamma a_{2}^{\dagger}a_{2}bc+\chi\Gamma a_{1}a_{1}^{\dagger}c^{\dagger}d+\chi\Gamma a_{2}^{\dagger}a_{2}c^{\dagger}d\\
 & + & \Gamma^{2}a_{1}a_{1}^{\dagger}a_{p}+\Gamma^{2}a_{2}^{\dagger}a_{2}a_{p}
\end{array}
\end{equation}
}{ }

{  
\begin{equation}
\begin{array}{lcl}
\left[G,\left[G,\left[G,a_{p}\left(0\right)\right]\right]\right] & = & -k_{p}^{3}a_{p}-\varGamma\left\{ k_{p}^{2}+k_{a_{1}}\left(k_{p}+k_{a_{1}}+k_{a_{2}}\right)\right.\\
 & + & \left.k_{a_{2}}\left(k_{p}+k_{a_{1}}+k_{a_{2}}\right)\right\} a_{1}a_{2}\\
 & - & g\varGamma\left(k_{p}+k_{a_{1}}+k_{a_{2}}\right)a_{1}a_{1}^{\dagger}bc\\
 & - & g\varGamma\left(k_{p}+k_{a_{1}}+k_{a_{2}}\right)a_{2}^{\dagger}a_{2}bc\\
 & - & \chi\Gamma\left(k_{p}+k_{a_{1}}+k_{a_{2}}\right)a_{1}a_{1}^{\dagger}c^{\dagger}d\\
 & - & \chi\Gamma\left(k_{p}+k_{a_{1}}+k_{a_{2}}\right)a_{2}^{\dagger}a_{2}c^{\dagger}d\\
 & - & \Gamma^{2}\left(k_{p}+k_{a_{1}}+k_{a_{2}}\right)a_{1}a_{1}^{\dagger}a_{p}\\
 & - & \Gamma^{2}\left(k_{p}+k_{a_{1}}+k_{a_{2}}\right)a_{2}^{\dagger}a_{2}a_{p}
\end{array}
\end{equation}
etc., ignoring all the terms beyond the quadratic powers of the interaction
constants $g,\,\chi$, and $\Gamma$. We use all the functions of
operators occurring in the higher-order nested commutators with unknown
coefficients $f_{n}$. From the set of assumed solution for the spatial
evolution all the operators we obtain a set of coupled differential
equations of all the coefficients $\mathcal{F}_{y}$ using Eq. (\ref{eq:assumedsol})
in Eq. (\ref{eq:equationofmotion}), as }{ }

{  
\begin{equation}
\begin{array}{lcl}
\overset{.}{f_{1}}\left(z\right) & = & ik_{p}j_{1}\\
\overset{.}{f_{2}}\left(z\right) & = & i\left(k_{p}f_{2}+\Gamma g_{1}h_{1}\right)\\
\overset{.}{f_{3}}\left(z\right) & = & i\left(k_{p}f_{3}+\Gamma g_{1}h_{2}\right)\\
\overset{.}{f_{4}}\left(z\right) & = & i\left(k_{p}f_{4}+\Gamma g_{2}h_{1}\right)\\
\overset{.}{f_{5}}\left(z\right) & = & i\left(k_{p}f_{5}+\Gamma g_{1}h_{3}\right)\\
\overset{.}{f_{6}}\left(z\right) & = & i\left(k_{p}f_{6}+\Gamma g_{3}h_{1}\right)\\
\overset{.}{f_{7}}\left(z\right) & = & i\left(k_{p}f_{7}+\Gamma g_{1}h_{4}\right)\\
\overset{.}{f_{8}}\left(z\right) & = & i\left(k_{p}f_{8}+\Gamma g_{4}h_{1}\right)
\end{array}
\end{equation}
}{ }

{   The solution of coupled differential equations of $\mathcal{F}_{y}$
with initial conditions $\mathcal{F}_{y}=\delta_{y,1}$ at $t=0$,
gives us the analytical form of $f_{n}$ as 
\begin{equation}
\begin{array}{lcl}
f_{1} & = & e^{izk_{p}}\\
\frac{f_{2}}{f_{1}} & = & \frac{\Gamma\left(e^{iz\text{\ensuremath{\Delta}k}_{D}}-1\right)}{\text{\ensuremath{\Delta}k}_{D}}\\
\frac{f_{3}}{f_{1}} & = & \frac{f_{4}}{f_{1}}=-\frac{\Gamma g\left(\text{\ensuremath{\Delta}k}_{D}\left(-e^{iz\text{\ensuremath{\Delta}k}_{3}}\right)+\text{\ensuremath{\Delta}k}_{3}e^{iz\text{\ensuremath{\Delta}k}_{D}}-\text{\ensuremath{\Delta}k}_{S}\right)}{\text{\ensuremath{\Delta}k}_{D}\text{\ensuremath{\Delta}k}_{S}\text{\ensuremath{\Delta}k}_{3}}\\
\frac{f_{5}}{f_{1}} & = & \frac{f_{6}}{f_{1}}=\frac{\Gamma\chi\left(\text{\ensuremath{\Delta}k}_{4}e^{iz\text{\ensuremath{\Delta}k}_{D}}+\text{\ensuremath{\Delta}k}_{D}e^{iz\text{\ensuremath{\Delta}k}_{4}}-\text{\ensuremath{\Delta}k}_{A}\right)}{\text{\ensuremath{\Delta}k}_{A}\text{\ensuremath{\Delta}k}_{D}\text{\ensuremath{\Delta}k}_{4}}\\
\frac{f_{7}}{f_{1}} & = & \frac{f_{8}}{f_{1}}-\frac{\Gamma^{2}\left(1+iz\text{\ensuremath{\Delta}k}_{D}-e^{iz\text{\ensuremath{\Delta}k}_{D}}\right)e^{izk_{p}}}{\text{\ensuremath{\Delta}k}_{D}^{2}}
\end{array}
\end{equation}
In the similar manner, we obtain the other $\mathcal{F}_{y}$s which
are }{ }

{  
\begin{equation}
\begin{array}{lcl}
g_{1} & = & e^{izk_{a_{1}}}\\
\frac{g_{2}}{g_{1}} & = & -\frac{g\left(1-e^{iz\text{\ensuremath{\Delta}k}_{S}}\right)}{\text{\ensuremath{\Delta}k}_{S}}\\
\frac{g_{3}}{g_{1}} & = & \frac{\chi\left(1-e^{-iz\text{\ensuremath{\Delta}k}_{A}}\right)}{\text{\ensuremath{\Delta}k}_{A}}\\
\frac{g_{4}}{g_{1}} & = & \frac{\Gamma\left(1-e^{-iz\text{\ensuremath{\Delta}k}_{D}}\right)}{\text{\ensuremath{\Delta}k}_{D}}\\
\frac{g_{5}}{g_{1}} & = & -\frac{g\chi\left(\text{\ensuremath{\Delta}k}_{A}-\text{\ensuremath{\Delta}k}_{2}e^{iz\text{\ensuremath{\Delta}k}_{S}}+\text{\ensuremath{\Delta}k}_{S}e^{iz\text{\ensuremath{\Delta}k}_{2}}\right)}{\text{\ensuremath{\Delta}k}_{A}\text{\ensuremath{\Delta}k}_{S}\text{\ensuremath{\Delta}k}_{2}}\\
\frac{g_{6}}{g_{1}} & = & -\frac{g\chi\left(\text{\ensuremath{\Delta}k}_{S}+\text{\ensuremath{\Delta}k}_{A}e^{-iz\text{\ensuremath{\Delta}k}_{2}}-\text{\ensuremath{\Delta}k}_{2}e^{-iz\text{\ensuremath{\Delta}k}_{A}}\right)}{\text{\ensuremath{\Delta}k}_{A}\text{\ensuremath{\Delta}k}_{S}\text{\ensuremath{\Delta}k}_{2}}\\
\frac{g_{7}}{g_{1}} & = & -\frac{\Gamma g\left(\text{\ensuremath{\Delta}k}_{D}+\text{\ensuremath{\Delta}k}_{S}e^{iz\text{\ensuremath{\Delta}k}_{3}}-\text{\ensuremath{\Delta}k}_{3}e^{iz\text{\ensuremath{\Delta}k}_{S}}\right)}{\text{\ensuremath{\Delta}k}_{D}\text{\ensuremath{\Delta}k}_{S}\text{\ensuremath{\Delta}k}_{3}}\\
\frac{g_{8}}{g_{1}} & = & -\frac{\Gamma g\left(\text{\ensuremath{\Delta}k}_{S}+\text{\ensuremath{\Delta}k}_{D}e^{-iz\text{\ensuremath{\Delta}k}_{3}}-\text{\ensuremath{\Delta}k}_{3}e^{-iz\text{\ensuremath{\Delta}k}_{D}}\right)}{\text{\ensuremath{\Delta}k}_{D}\text{\ensuremath{\Delta}k}_{S}\text{\ensuremath{\Delta}k}_{3}}\\
\frac{g_{9}}{g_{1}} & = & -\frac{\Gamma\chi\left(\text{\ensuremath{\Delta}k}_{D}-\text{\ensuremath{\Delta}k}_{A}e^{-iz\text{\ensuremath{\Delta}k}_{4}}+\text{\ensuremath{\Delta}k}_{4}e^{-iz\text{\ensuremath{\Delta}k}_{A}}\right)}{\text{\ensuremath{\Delta}k}_{A}\text{\ensuremath{\Delta}k}_{D}\text{\ensuremath{\Delta}k}_{4}}\\
\frac{g_{10}}{g_{1}} & = & \frac{\Gamma\chi\left(\text{\ensuremath{\Delta}k}_{A}-\text{\ensuremath{\Delta}k}_{D}e^{iz\text{\ensuremath{\Delta}k}_{4}}-\text{\ensuremath{\Delta}k}e^{-iz\text{\ensuremath{\Delta}k}_{D}}\right)}{\text{\ensuremath{\Delta}k}_{A}\text{\ensuremath{\Delta}k}_{D}\text{\ensuremath{\Delta}k}_{4}}\\
\frac{g_{11}}{g_{1}} & = & -\frac{g_{12}}{g_{1}}=-\frac{g_{13}}{g_{1}}=\frac{g^{2}\left(1+iz\text{\ensuremath{\Delta}k}_{S}-e^{iz\text{\ensuremath{\Delta}k}_{S}}\right)}{\text{\ensuremath{\Delta}k}_{S}^{2}}\\
\frac{g_{14}}{g_{1}} & = & \frac{g_{15}}{g_{1}}=-\frac{g_{17}}{g_{1}}=\frac{\chi^{2}\left(1-iz\text{\ensuremath{\Delta}k}_{A}-e^{-z\text{i\ensuremath{\Delta}k}_{A}}\right)}{\text{\ensuremath{\Delta}k}_{A}^{2}}\\
\frac{g_{16}}{g_{1}} & = & -\frac{g_{18}}{g_{1}}=\frac{\Gamma^{2}\left(1-iz\text{\ensuremath{\Delta}k}_{D}-e^{-iz\text{\ensuremath{\Delta}k}_{D}}\right)}{\text{\ensuremath{\Delta}k}_{D}^{2}}\\
\frac{g_{19}}{g_{1}} & = & -\frac{g\chi\left(\text{\ensuremath{\Delta}k}_{S}+\text{\ensuremath{\Delta}k}_{A}-\text{\ensuremath{\Delta}k}_{2}e^{-iz\text{\ensuremath{\Delta}k}_{1}}+\text{\ensuremath{\Delta}k}_{1}\left(e^{-iz\text{\ensuremath{\Delta}k}_{A}}-e^{iz\text{\ensuremath{\Delta}k}_{S}}\right)\right)}{\text{\ensuremath{\Delta}k}_{A}\text{\ensuremath{\Delta}k}_{S}\text{\ensuremath{\Delta}k}_{1}}
\end{array}
\end{equation}
}{ }

{  
\begin{equation}
\begin{array}{lcl}
h_{1} & = & e^{izk_{a_{2}}}\\
\frac{h_{2}}{h_{1}} & = & -\frac{g\left(1-e^{iz\text{\ensuremath{\Delta}k}_{S}}\right)}{\text{\ensuremath{\Delta}k}_{S}}\\
\frac{h_{3}}{h_{1}} & = & \frac{\chi\left(1-e^{-iz\text{\ensuremath{\Delta}k}_{A}}\right)}{\text{\ensuremath{\Delta}k}_{A}}\\
\frac{h_{4}}{h_{1}} & = & \frac{\Gamma\left(1-e^{-iz\text{\ensuremath{\Delta}k}_{D}}\right)}{\text{\ensuremath{\Delta}k}_{D}}\\
\frac{h_{5}}{h_{1}} & = & -\frac{g\chi\left(\text{\ensuremath{\Delta}k}_{A}-\text{\ensuremath{\Delta}k}_{2}e^{iz\text{\ensuremath{\Delta}k}_{S}}+\text{\ensuremath{\Delta}k}_{S}e^{iz\text{\ensuremath{\Delta}k}_{2}}\right)}{\text{\ensuremath{\Delta}k}_{A}\text{\ensuremath{\Delta}k}_{S}\text{\ensuremath{\Delta}k}_{2}}\\
\frac{h_{6}}{h_{1}} & = & -\frac{g\chi\left(\text{\ensuremath{\Delta}k}_{S}+\text{\ensuremath{\Delta}k}_{A}e^{-iz\left(\text{\ensuremath{\Delta}k}_{A}+\text{\ensuremath{\Delta}k}_{S}\right)}-\text{\ensuremath{\Delta}k}_{2}e^{-iz\text{\ensuremath{\Delta}k}_{A}}\right)}{\text{\ensuremath{\Delta}k}_{A}\text{\ensuremath{\Delta}k}_{S}\text{\ensuremath{\Delta}k}_{2}}\\
\frac{h_{7}}{h_{1}} & = & -\frac{\Gamma ge^{izk_{a_{1}}}\left(\text{\ensuremath{\Delta}k}_{D}+\text{\ensuremath{\Delta}k}_{S}e^{iz\left(\text{\ensuremath{\Delta}k}_{D}+\text{\ensuremath{\Delta}k}_{S}\right)}-\left(\text{\ensuremath{\Delta}k}_{D}+\text{\ensuremath{\Delta}k}_{S}\right)e^{iz\text{\ensuremath{\Delta}k}_{S}}\right)}{\text{\ensuremath{\Delta}k}_{D}\text{\ensuremath{\Delta}k}_{S}\left(\text{\ensuremath{\Delta}k}_{D}+\text{\ensuremath{\Delta}k}_{S}\right)}\\
\frac{h_{8}}{h_{1}} & = & -\frac{\Gamma g\left(\text{\ensuremath{\Delta}k}_{S}+\text{\ensuremath{\Delta}k}_{D}e^{-iz\left(\text{\ensuremath{\Delta}k}_{D}+\text{\ensuremath{\Delta}k}_{S}\right)}-\left(\text{\ensuremath{\Delta}k}_{D}+\text{\ensuremath{\Delta}k}_{S}\right)e^{-iz\text{\ensuremath{\Delta}k}_{D}}\right)}{\text{\ensuremath{\Delta}k}_{D}\text{\ensuremath{\Delta}k}_{S}\left(\text{\ensuremath{\Delta}k}_{D}+\text{\ensuremath{\Delta}k}_{S}\right)}\\
\frac{h_{9}}{h_{1}} & = & -\frac{\Gamma\chi\left(\text{\ensuremath{\Delta}k}_{D}+\text{\ensuremath{\Delta}k}_{A}\left(-e^{iz\left(\text{\ensuremath{\Delta}k}_{D}-\text{\ensuremath{\Delta}k}_{A}\right)}\right)+\left(\text{\ensuremath{\Delta}k}_{A}-\text{\ensuremath{\Delta}k}_{D}\right)e^{-iz\text{\ensuremath{\Delta}k}_{A}}\right)}{\text{\ensuremath{\Delta}k}_{A}\text{\ensuremath{\Delta}k}_{D}\left(\text{\ensuremath{\Delta}k}_{A}-\text{\ensuremath{\Delta}k}_{D}\right)}\\
\frac{h_{10}}{h_{1}} & = & \frac{\Gamma\chi\left(\text{\ensuremath{\Delta}k}_{A}-\text{\ensuremath{\Delta}k}_{D}e^{iz\left(\text{\ensuremath{\Delta}k}_{A}-\text{\ensuremath{\Delta}k}_{D}\right)}+\left(\text{\ensuremath{\Delta}k}_{D}-\text{\ensuremath{\Delta}k}_{A}\right)e^{-iz\text{\ensuremath{\Delta}k}_{D}}\right)}{\text{\ensuremath{\Delta}k}_{A}\text{\ensuremath{\Delta}k}_{D}\left(\text{\ensuremath{\Delta}k}_{A}-\text{\ensuremath{\Delta}k}_{D}\right)}\\
\frac{h_{11}}{h_{1}} & = & -\frac{h_{12}}{h_{1}}=-\frac{h_{13}}{h_{1}}=\frac{g^{2}\left(1+iz\text{\ensuremath{\Delta}k}_{S}-e^{iz\text{\ensuremath{\Delta}k}_{S}}\right)}{\text{\ensuremath{\Delta}k}_{S}^{2}}\\
\frac{h_{14}}{h_{1}} & = & \frac{h_{15}}{h_{1}}=-\frac{h_{17}}{h_{1}}=\frac{\chi^{2}\left(1-iz\text{\ensuremath{\Delta}k}_{A}-e^{-z\text{i\ensuremath{\Delta}k}_{A}}\right)}{\text{\ensuremath{\Delta}k}_{A}^{2}}\\
\frac{h_{16}}{h_{1}} & = & -\frac{h_{18}}{h_{1}}=\frac{\Gamma^{2}\left(1-iz\text{\ensuremath{\Delta}k}_{D}-e^{-iz\text{\ensuremath{\Delta}k}_{D}}\right)}{\text{\ensuremath{\Delta}k}_{D}^{2}}\\
\frac{h_{19}}{h_{1}} & = & -\frac{g\chi e^{izk_{a_{1}}}\left(\text{\ensuremath{\Delta}k}_{S}+\text{\ensuremath{\Delta}k}_{A}-\left(\text{\ensuremath{\Delta}k}_{A}+\text{\ensuremath{\Delta}k}_{S}\right)e^{-iz\left(\text{\ensuremath{\Delta}k}_{A}-\text{\ensuremath{\Delta}k}_{S}\right)}+\left(\text{\ensuremath{\Delta}k}_{A}-\text{\ensuremath{\Delta}k}_{S}\right)\left(e^{-iz\text{\ensuremath{\Delta}k}_{A}}-e^{iz\text{\ensuremath{\Delta}k}_{S}}\right)\right)}{\text{\ensuremath{\Delta}k}_{A}\text{\ensuremath{\Delta}k}_{S}\left(\text{\ensuremath{\Delta}k}_{A}-\text{\ensuremath{\Delta}k}_{S}\right)}
\end{array}
\end{equation}
}{ }

{  
\begin{equation}
\begin{array}{lcl}
j_{1} & = & e^{iz{k}_{b}},\\
\frac{j_{2}}{j_{1}} & = & \frac{g\left(1-e^{-iz\Delta{k}_{S}}\right)}{\Delta{k}_{S}},\\
\frac{j_{3}}{j_{1}} & = & \frac{g\chi\left(\Delta{k}_{A}-\Delta{k}_{1}e^{-iz\Delta{k}_{S}}-\Delta{k}_{S}e^{iz\Delta{k}_{1}}\right)}{\Delta{k}_{A}\Delta{k}_{1}\Delta{k}_{S}},\\
\frac{j_{4}}{j_{1}} & = & \frac{j_{5}}{j_{1}}=\frac{g\chi\left(\Delta{k}_{A}+\Delta{k}_{S}e^{-iz\Delta{k}_{2}}-\Delta{k}_{2}e^{-iz\Delta{k}_{S}}\right)}{\Delta{k}_{A}\Delta{k}_{S}\Delta{k}_{2}},\\
\frac{j_{6}}{j_{1}} & = & \frac{j_{7}}{j_{1}}=\frac{g\Gamma\left(\Delta{k}_{D}+\Delta{k}_{S}e^{-iz\Delta{k}_{3}}-\Delta{k}_{3}e^{-iz\Delta{k}_{S}}\right)}{\Delta{k}_{D}\Delta{k}_{S}\Delta{k}_{3}},\\
\frac{j_{8}}{j_{1}} & = & -\frac{j_{9}}{j_{1}}=-\frac{j_{10}}{j_{1}}=\frac{g^{2}\left(1-e^{-iz\Delta{k}_{S}}-i\Delta{k}_{S}z\right)}{\Delta{k}_{S}^{2}},\label{eq:js}
\end{array}
\end{equation}
\begin{equation}
{\normalcolor {\color{blue}{\normalcolor {\color{blue}{\normalcolor \begin{array}{lcl}
k_{1} & = & e^{iz{k}_{c}},\\
\frac{k_{2}}{k_{1}} & = & \frac{g\left(1-e^{-iz\Delta{k}_{S}}\right)}{\Delta{k}_{S}},\\
\frac{k_{3}}{k_{1}} & = & \frac{\chi\left(1-e^{-iz\Delta{k}_{A}}\right)}{\Delta{k}_{A}},\\
\frac{k_{4}}{k_{1}} & = & \frac{k_{5}}{k_{1}}=\frac{g\Gamma\left(\Delta{k}_{D}+\Delta{k}_{S}e^{-iz\Delta{k}_{3}}-\Delta{k}_{3}e^{-iz\Delta{k}_{S}}\right)}{\Delta{k}_{D}\Delta{k}_{S}\Delta{k}_{3}},\\
\frac{k_{6}}{k_{1}} & = & \frac{k_{7}}{k_{1}}=\frac{\Gamma\chi\left(-\Delta{k}_{D}+\Delta{k}_{A}e^{-iz\Delta{k}_{4}}-\Delta{k}_{4}e^{-iz\Delta{k}_{A}}\right)}{\Delta{k}_{A}\Delta{k}_{D}\Delta{k}_{4}},\\
\frac{k_{8}}{k_{1}} & = & -\frac{k_{11}}{k_{1}}=-\frac{k_{12}}{k_{1}}=-\frac{\chi^{2}\left(1-e^{-iz\Delta{k}_{A}}-i\Delta{k}_{A}z\right)}{\Delta{k}_{A}^{2}},\\
\frac{k_{9}}{k_{1}} & = & \frac{k_{10}}{k_{1}}=-\frac{k_{13}}{k_{1}}=-\frac{g^{2}\left(1-e^{iz\Delta{k}_{S}}-i\Delta{k}_{S}z\right)}{\Delta{k}_{S}^{2}},\\
\frac{k_{14}}{k_{1}} & = & \frac{k_{15}}{k_{1}}=\frac{g\chi\left\{ \Delta{k}_{2}\left(e^{-i\Delta{k}_{A}z}-e^{-i\Delta{k}_{S}z}\right)+\Delta{k}_{1}\left(1-e^{-i\Delta{k}_{2}z}\right)\right\} }{\Delta{k}_{S}\Delta{k}_{A}\Delta{k}_{2}},
\end{array}}}}}}
\end{equation}
\begin{equation}
\begin{array}{lcl}
l_{1} & = & e^{iz{k}_{d}},\\
\frac{l_{2}}{l_{1}} & = & -\frac{\chi\left(1-e^{iz\Delta{k}_{A}}\right)}{\Delta{k}_{A}},\\
\frac{l_{3}}{l_{1}} & = & \frac{l_{4}}{l_{1}}=\frac{g\chi\left(\Delta{k}_{S}+\Delta{k}_{A}e^{iz\Delta{k}_{2}}-\Delta{k}_{2}e^{iz\Delta{k}_{A}}\right)}{\Delta{k}_{S}\Delta{k}_{A}\Delta{k}_{2}},\\
\frac{l_{5}}{l_{1}} & = & \frac{g\chi\left(\Delta{k}_{S}+\Delta{k}_{1}e^{iz\Delta{k}_{A}}-\Delta{k}_{A}e^{iz\Delta{k}_{1}}\right)}{\Delta{k}_{S}\Delta{k}_{1}\Delta{k}_{A}},\\
\frac{l_{6}}{l_{1}} & = & \frac{l_{7}}{l_{1}}=\frac{\Gamma\chi\left(\Delta{k}_{D}-\Delta{k}_{A}e^{iz\Delta{k}_{4}}+\Delta{k}_{4}e^{iz\Delta{k}_{A}}\right)}{\Delta{k}_{D}\Delta{k}_{A}\Delta{k}_{4}},\\
\frac{l_{8}}{l_{1}} & = & \frac{l_{9}}{l_{1}}=\frac{l_{10}}{l_{1}}=-\frac{\chi^{2}\left(1-e^{iz\Delta{k}_{A}}+i\Delta{k}_{A}z\right)}{\Delta{k}_{A}^{2}},
\end{array}
\end{equation}
where we use $\Delta{k}_{S}=\left(-{k}_{a_{1}}-{k}_{a_{2}}+{k}_{b}+{k}_{c}\right),$
$\Delta{k}_{A}=\left({k}_{a_{1}}+{k}_{a_{2}}+{k}_{c}-{k}_{d}\right)$,
$\Delta{k}_{D}=\left({k}_{a_{1}}+{k}_{a_{2}}-{k}_{p}\right)$, $\Delta{k}_{1}=\Delta{k}_{A}-\Delta{k}_{S}$,
$\Delta{k}_{2}=\Delta{k}_{A}+\Delta{k}_{S}$, $\Delta{k}_{3}=\Delta{k}_{S}+\Delta{k}_{D}$,
and $\Delta{k}_{4}=\Delta{k}_{A}-\Delta{k}_{D}.$ Here, it will be
apt to note that in a strict sense, the validity of the short-length
solution \citep{val-ST} is limited by $\left|\Lambda z\xi\right|<1$,
where $\Lambda=\max\left\{ g,\chi,\Gamma\right\} $ and $\xi=\max\xi\,\forall\xi\in\left\{ \alpha,\alpha_{1},\alpha_{2},\beta,\gamma,\delta\right\} $.
This gives us the validity of the short-length solution for propagation
lengths $z<1/\left|\Lambda\xi\right|$, while the present perturbative
solution has enlarged validity for finite times due to modulation
by the sinc function related to the phase mismatches, which ensures
the convergence (see \citep{off-res-Raman} for detail).}{ }

\section*{   Appendix B: Nonclassicality Quantifiers\protect\label{sec:app-B}}

{   Utilising equations (\ref{eq:inSt}), (\ref{eq:steering-criteria}), 
and (\ref{eq:assumedsol}), we derive the expression for quantum steering
over the various combined modes, detailed as follows:}{ }

\begin{equation}
\begin{array}{lcl}
S_{a_{1}\rightarrow b} & = & \frac{\left|\alpha_{1}\right|^{2}}{2} + \left|g_{2}\right|^{2} \left[ \left|\alpha_{1}\right|^{4} \left|\alpha_{2}\right|^{2} + \left|\beta\right|^{4} \left|\gamma\right|^{2} 
- \frac{1}{2} \left( \left|\alpha_{1}\right|^{2} - \left|\alpha_{2}\right|^{2} - 1 \right) \left|\beta\right|^{2} \left|\gamma\right|^{2} \right. \\
 & - & \left. \frac{1}{2} \left|\alpha_{1}\right|^{2} \left|\alpha_{2}\right|^{2} \left( 3 \left|\beta\right|^{2} + \left|\gamma\right|^{2} + 1 \right) \right] \\
 & + & \left|g_{3}\right|^{2} \left[ \left( \left|\alpha_{2}\right|^{2} - \left|\gamma\right|^{2} \right) \left|\beta\right|^{2} \left|\delta\right|^{2} 
+ \left( \left|\alpha_{2}\right|^{2} + 1 \right) \left( \left|\gamma\right|^{2} + 1 \right) \frac{\left|\delta\right|^{2}}{2} \right. \\
 & + & \left. \frac{1}{2} \left|\alpha_{1}\right|^{2} \left|\delta\right|^{2} \left( \left|\alpha_{2}\right|^{2} + \left|\gamma\right|^{2} + 1 \right) 
- \frac{\left|\alpha_{1}\right|^{2} \left|\alpha_{2}\right|^{2} \left|\gamma\right|^{2}}{2} \right] \\
 & + & \left|g_{4}\right|^{2} \left[ \frac{\left|\alpha\right|^{2}}{2} \left( 2 \left|\beta\right|^{2} + \left|\alpha_{2}\right|^{2} + \left|\alpha_{1}\right|^{2} + 1 \right) 
- \frac{\left|\alpha_{1}\right|^{2} \left|\alpha_{2}\right|^{2}}{2} \right] \\
 & + & \left[ \left\{ \frac{g_{1} g_{2}^{*}}{2} \alpha_{1} \alpha_{2} \beta^{*} \gamma^{*} + \frac{g_{1} g_{3}^{*}}{2} \alpha_{1} \alpha_{2} \gamma \delta^{*} \right. \right. \\
 & + & \frac{g_{1} g_{4}^{*}}{2} \alpha^{*} \alpha_{1} \alpha_{2} 
+ \left( \frac{g_{1}^{*} g_{6} - g_{1} g_{5}^{*}}{2} \right) \left|\alpha_{1}\right|^{2} \beta^{*} \gamma^{2*} \delta \\
 & + & \left( \frac{g_{1}^{*} g_{8} - g_{1} g_{7}^{*}}{2} \right) \left|\alpha_{1}\right|^{2} \alpha \beta^{*} \gamma^{*} 
+ \left( \frac{g_{1}^{*} g_{9} + g_{1} g_{10}^{*}}{2} \right) \left|\alpha_{1}\right|^{2} \alpha \gamma \delta^{*} \\
 & + & \left. \left. \left( \left|\beta\right|^{2} + \frac{\left|\alpha_{2}\right|^{2} + 1}{2} \right) 
\left( g_{2} g_{3}^{*} \beta \gamma^{2} \delta^{*} + g_{2} g_{4}^{*} \alpha^{*} \beta \gamma + g_{3} g_{4}^{*} \alpha^{*} \gamma^{*} \delta \right) \right\} + {\rm c.c.} \right]
\end{array}
\end{equation}

\begin{equation}
\begin{array}{lcl}
S_{a_{1}\rightarrow c} & = & \frac{\left|\alpha_{1}\right|^{2}}{2} + \left|g_{2}\right|^{2} \left[ \frac{\left|\alpha_{1}\right|^{2} \left|\alpha_{2}\right|^{2}}{2} \left( 2 \left|\alpha_{1}\right|^{2} 
- \left|\beta\right|^{2} - 3 \left|\gamma\right|^{2} - 1 \right) \right. \\
 & - & \left. \frac{1}{2} \left( \left|\alpha_{1}\right|^{2} - \left|\alpha_{2}\right|^{2} 
- 2 \left|\gamma\right|^{2} - 1 \right) \left|\beta\right|^{2} \left|\gamma\right|^{2} \right] \\
 & + & \left|g_{3}\right|^{2} \left[ \left( \left|\alpha_{1}\right|^{4} + \left|\gamma\right|^{4} \right) \left|\delta\right|^{2} + 3 \left|\delta\right|^{2} 
\left( \left|\alpha_{2}\right|^{2} + 1 \right) \left( \left|\alpha_{1}\right|^{2} + \left|\gamma\right|^{2} \right) \right. \\
 & + & \left. \frac{\left|\alpha_{1}\right|^{2} \left|\delta\right|^{2}}{2} \left( \left|\alpha_{2}\right|^{2} + 3 \left|\gamma\right|^{2} + 1 \right) \right. \\
 & + & \left. \left|\delta\right|^{2} \left( \left|\alpha_{2}\right|^{2} + \frac{\left(\left|\alpha_{2}\right|^{2} + 1 \right) 
\left(\left|\gamma\right|^{2} + 1 \right)}{2} \right) - \frac{3}{2} \left|\alpha_{1}\right|^{2} \left|\alpha_{2}\right|^{2} \left|\gamma\right|^{2} \right] \\
 & + & \left|g_{4}\right|^{2} \left[ \left|\alpha\right|^{2} \left( \left|\gamma\right|^{2} 
+ \frac{\left|\alpha_{1}\right|^{2} + \left|\alpha_{2}\right|^{2} + 1}{2} \right) - \frac{\left|\alpha_{1}\right|^{2} \left|\alpha_{2}\right|^{2}}{2} \right] \\
 & + & \left[ \left\{ \frac{g_{1} g_{2}^{*}}{2} \alpha_{1} \alpha_{2} \beta^{*} \gamma^{*} 
+ \frac{3}{2} g_{1} g_{3}^{*} \alpha_{1} \alpha_{2} \gamma \delta^{*} \right. \right. \\
 & + & \frac{g_{1} g_{4}^{*}}{2} \alpha^{*} \alpha_{1} \alpha_{2} 
+ \frac{ \left( g_{1} g_{5}^{*} + 5 g_{1} g_{6}^{*} \right)}{2} \left|\alpha_{1}\right|^{2} \beta \gamma^{2} \delta^{*} \\
 & + & \frac{\left( g_{1}^{*} g_{8} - g_{1} g_{7}^{*} \right)}{2} \alpha \left|\alpha_{1}\right|^{2} \beta^{*} \gamma^{*} 
+ \left( \frac{g_{1}^{*} g_{10} + 3 g_{1} g_{9}^{*}}{2} \right) \alpha \left|\alpha_{1}\right|^{2} \gamma \delta^{*} \\
 & + & g_{2} g_{3}^{*} \left( - \left|\alpha_{1}\right|^{2} + \left|\alpha_{2}\right|^{2} + \left|\gamma\right|^{2} + 1 
+ \frac{\left|\alpha_{2}\right|^{2} + 1}{2} \right) \beta \gamma^{2} \delta^{*} \\
 & + & \left( \frac{\left|\alpha_{2}\right|^{2} + 1}{2} \right) \left( g_{2} g_{4}^{*} \alpha^{*} \beta \gamma 
+ 3 g_{3} g_{4}^{*} \alpha^{*} \gamma^{*} \delta \right) \\
 & + & \left. \left. \left|\gamma\right|^{2} \left( g_{2} g_{4}^{*} \alpha^{*} \beta \gamma 
+ g_{3} g_{4}^{*} \alpha^{*} \gamma^{*} \delta \right) 
+ \left( \frac{g_{1} g_{19}^{*} + g_{1}^{*} g_{19}}{2} - k_{2} k_{3}^{*} \right) \alpha_{1}^{2} \alpha_{2}^{2} \beta^{*} \delta^{*} \right\} + {\rm c.c.} \right]
\end{array}
\end{equation}

\begin{equation}
\begin{array}{lcl}
S_{a_{1} \rightarrow d} & = & \left|g_{2}\right|^{2} \left[ \left( \left|\delta\right|^{2} 
+ \frac{\left|\alpha_{1}\right|^{2} + \left|\alpha_{2}\right|^{2} + 1}{2} \right) 
\left|\beta\right|^{2} \left|\gamma\right|^{2} \right. \\
 & - & \left. \frac{\left|\alpha_{1}\right|^{2} \left|\alpha_{2}\right|^{2}}{2} 
\left( \left|\beta\right|^{2} + \left|\gamma\right|^{2} + 1 \right) \right] \\
 & + & \left|g_{3}\right|^{2} \left[ -\frac{\left|\alpha_{1}\right|^{2} \left|\delta\right|^{2}}{2} 
\left( \left|\alpha_{2}\right|^{2} + \left|\gamma\right|^{2} + 1 \right) \right. \\
 & + & \left. \frac{\left|\delta\right|^{2}}{2} \left( \left|\alpha_{2}\right|^{2} + \left|\gamma\right|^{2} + 1 \right) 
+ \left|\delta\right|^{4} \left( \left|\alpha_{2}\right|^{2} + \left|\gamma\right|^{2} + 1 \right) \right. \\
 & + & \left. \frac{\left|\alpha_{2}\right|^{2} \left|\gamma\right|^{2}}{2} 
\left( \left|\alpha_{1}\right|^{2} + \left|\delta\right|^{2} \right) \right] \\
 & + & \left|g_{4}\right|^{2} \left[ \left|\alpha\right|^{2} \left|\delta\right|^{2} 
+ \frac{\left|\alpha\right|^{2}}{2} \left( \left|\alpha_{2}\right|^{2} + \left|\alpha_{1}\right|^{2} + 1 \right) 
- \frac{\left|\alpha_{1}\right|^{2} \left|\alpha_{2}\right|^{2}}{2} \right] \\
 & + & \frac{\left|\alpha_{1}\right|^{2}}{2} \\
 & + & \left[ \left\{ \frac{g_{1} g_{2}^{*}}{2} \alpha_{1} \alpha_{2} \beta^{*} \gamma^{*} 
+ \frac{g_{1} g_{3}^{*}}{2} \alpha_{1} \alpha_{2} \gamma \delta^{*} 
+ \frac{g_{1} g_{4}^{*}}{2} \alpha^{*} \alpha_{1} \alpha_{2} \right. \right. \\
 & + & \left( \frac{g_{1}^{*} g_{5} - g_{1} g_{6}^{*}}{2} \right) \left|\alpha_{1}\right|^{2} \beta \gamma^{2} \delta^{*} \\
 & + & \left( \frac{g_{1} g_{7}^{*} + g_{1}^{*} g_{8}}{2} \right) \alpha \left|\alpha_{1}\right|^{2} \beta^{*} \gamma^{*} \\
 & + & \left( \frac{g_{1}^{*} g_{10} - g_{1} g_{9}^{*}}{2} \right) \alpha \left|\alpha_{1}\right|^{2} \gamma \delta^{*} 
+ \frac{g_{1} g_{19}^{*}}{2} \alpha_{1}^{2} \alpha_{2}^{2} \beta^{*} \delta^{*} \\
 & + & \left. \left. \left( \left|\delta\right|^{2} + \frac{\left|\alpha_{2}\right|^{2} + 1}{2} \right) 
\left( g_{2} g_{3}^{*} \beta \gamma^{2} \delta^{*} + g_{2} g_{4}^{*} \alpha^{*} \beta \gamma 
+ g_{3} g_{4}^{*} \alpha^{*} \gamma^{*} \delta \right) \right\} + {\rm c.c.}. \right]
\end{array}
\end{equation}

\begin{equation}
\begin{array}{lcl}
S_{bc} & = & \frac{\left|\beta\right|^{2}}{2} + \frac{\left|j_{2}\right|^{2}}{2} \left[ \left|\alpha_{1}\right|^{2} \left|\alpha_{2}\right|^{2} \left( 7\left|\beta\right|^{2} + 7\left|\gamma\right|^{2} + 3 \right) \right. \\
 & - & \left. 3\left|\beta\right|^{2} \left|\gamma\right|^{2} \left( \left|\alpha_{1}\right|^{2} + \left|\alpha_{2}\right|^{2} + 1 \right) \right] \\
 & + & \left[ \left\{ \frac{3}{2} j_{1} j_{2}^{\ast} \alpha_{1}^{\ast} \alpha_{2}^{\ast} \beta \gamma 
+ \frac{3}{2} j_{1} j_{6}^{\ast} \left( \left|\alpha_{1}\right|^{2} + \left|\alpha_{2}\right|^{2} + 1 \right) \alpha^{*} \beta \gamma \right. \right. \\
 & + & \left. \frac{5}{2} j_{1} j_{4}^{\ast} \left( \left|\alpha_{1}\right|^{2} + \left|\alpha_{2}\right|^{2} + 1 \right) \beta \gamma^{2} \delta^{\ast} \right. \\
 & + & \left. \left. \left( k_{2} k_{3}^{\ast} + \frac{j_{1}^{\ast} j_{3}}{2} \right) \alpha_{1}^{2} \alpha_{2}^{2} \beta^{*} \delta^{*} \right\} + {\rm c.c.} \right]
\end{array}
\end{equation}

{  
\begin{equation}
\begin{array}{lcl}
S_{b\rightarrow d} & = & \frac{\left|\beta\right|^{2}}{2}+\left|j_{2}\right|^{2}\left[\left|\alpha_{1}\right|^{2}\left|\alpha_{2}\right|^{2}\left(\left|\delta\right|^{2}+\frac{\left(\left|\beta\right|^{2}+\left|\gamma\right|^{2}+1\right)}{2}\right)-\frac{1}{2}\left(\left|\alpha_{1}\right|^{2}+\left|\alpha_{2}\right|^{2}+1\right)\left|\beta\right|^{2}\left|\gamma\right|^{2}\right]\\
 & + & \left[\left\{ \left(l_{1}^{*}l_{5}+\frac{j_{1}^{\ast}j_{3}}{2}\right)\alpha_{1}^{2}\alpha_{2}^{2}\beta^{*}\delta^{*}+\frac{j_{1}^{*}j_{2}}{2}\alpha_{1}\alpha_{2}\beta^{*}\gamma^{*}+\frac{j_{1}j_{4}^{*}}{2}\left(\left|\alpha_{1}\right|^{2}+\left|\alpha_{2}\right|^{2}+1\right)\beta\gamma^{2}\delta^{\ast}\right.\right.\\
 & + & \left.\left.\frac{j_{1}j_{6}^{\ast}}{2}\left(\left|\alpha_{1}\right|^{2}+\left|\alpha_{2}\right|^{2}+1\right)\alpha^{*}\beta\gamma\right\} +{\rm c.c.}\right]
\end{array}
\end{equation}
}{ }

\begin{equation}
\begin{array}{lcl}
S_{c\rightarrow d} & = & \frac{\left|\gamma\right|^{2}}{2} + \left|k_{2}\right|^{2} \left[ \left|\alpha_{1}\right|^{2} \left|\alpha_{2}\right|^{2} \left( \left|\beta\right|^{2} + \left|\delta\right|^{2} + 1 + \frac{\left|\gamma\right|^{2}}{2} \right) \right.  -  \left. \frac{1}{2} \left( \left|\alpha_{1}\right|^{2} + \left|\alpha_{2}\right|^{2} + 1 \right) \left|\beta\right|^{2} \left|\gamma\right|^{2} \right] \\
 & + & \left|k_{3}\right|^{2} \left[ \frac{\left|\alpha_{1}\right|^{2} \left|\alpha_{2}\right|^{2} \left( \left|\delta\right|^{2} - \left|\gamma\right|^{2} \right)}{2} + \left( \left|\alpha_{1}\right|^{2} + \left|\alpha_{2}\right|^{2} + 1 \right) \left|\delta\right|^{2} \left( \left|\delta\right|^{2} + \frac{3}{2} \left|\gamma\right|^{2} + \frac{5}{2} \right) \right] \\
 & + & \left[ \left\{ \frac{k_{1} k_{2}^{\ast}}{2} \alpha_{1}^{\ast} \alpha_{2}^{\ast} \beta \gamma + \frac{k_{1} k_{3}^{\ast}}{2} \alpha_{1} \alpha_{2} \gamma \delta^{*} + \frac{k_{1} k_{4}^{\ast}}{2} \left( \left|\alpha_{1}\right|^{2} + \left|\alpha_{2}\right|^{2} + 1 \right) \alpha^{*} \beta \gamma \right. \right. \\
 & + & \left. \frac{k_{1} k_{6}^{\ast}}{2} \left( \left|\alpha_{1}\right|^{2} + \left|\alpha_{2}\right|^{2} + 1 \right) \alpha \gamma \delta^{*} + \frac{k_{1} k_{14}^{\ast}}{2} \left( \left|\alpha_{1}\right|^{2} + \left|\alpha_{2}\right|^{2} + 1 \right) \beta \gamma^{2} \delta^{\ast} \right. \\
 & + & \left. \left. k_{2} k_{3}^{\ast} \alpha_{1}^{2} \alpha_{2}^{2} \beta^{*} \delta^{*} \right\} + {\rm c.c.} \right]
\end{array}
\end{equation}

\begin{equation}
\begin{array}{lcl}
S_{d\rightarrow a_{1}} & = & \left[\frac{\left|\delta\right|^{2}}{2} + \left|g_{2}\right|^{2} \left|\beta\right|^{2} \left|\gamma\right|^{2} \left|\delta\right|^{2} + \left|g_{3}\right|^{2} \left\{ \left( \left|\alpha_{2}\right|^{2} + \left|\gamma\right|^{2} + 1 \right) \left|\delta\right|^{2} \right. \right. \\
 & \times & \left( \left|\delta\right|^{2} - \frac{1}{2} \right) - \frac{3}{2} \left|\alpha_{1}\right|^{2} \left|\delta\right|^{2} \left( \left|\alpha_{2}\right|^{2} + 1 \right) \\
 & - & \left. \frac{\left|\gamma\right|^{2} \left|\delta\right|^{2}}{2} \left( 3 \left|\alpha_{1}\right|^{2} + \left|\alpha_{2}\right|^{2} \right) + \frac{\left|\alpha_{1}\right|^{2} \left|\alpha_{2}\right|^{2} \left|\gamma\right|^{2}}{2} \right\} \\
 & + & \left|g_{4}\right|^{2} \left|\alpha\right|^{2} \left|\delta\right|^{2} + \left\{ \left(-g_{1} g_{6}^{*} + \frac{l_{1}^{\star} l_{3}}{2} \right) \left|\alpha_{1}\right|^{2} \beta \gamma^{2} \delta^{*} \right. \\
 & + & \left(-g_{1} g_{9}^{*} + \frac{l_{1}^{\star} l_{6}}{2} \right) \left|\alpha_{1}\right|^{2} \alpha \gamma \delta^{*} + g_{2} g_{3}^{*} \left|\delta\right|^{2} \beta \gamma^{2} \delta^{*} \\
 & + & g_{2} g_{4}^{*} \alpha^{*} \beta \gamma \left|\delta\right|^{2} + g_{3} g_{4}^{*} \left|\delta\right|^{2} \alpha^{*} \gamma^{*} \delta + \frac{l_{1}^{\star} l_{2} \alpha_{1} \alpha_{2} \gamma \delta^{\ast}}{2} \\
 & + & \frac{l_{1}^{\star} l_{3} \beta \gamma^{2} \delta^{\ast}}{2} + \frac{l_{1}^{\star} l_{4} \left|\alpha_{2}\right|^{2} \beta \gamma^{2} \delta^{\ast}}{2} + \frac{l_{1}^{\star} l_{5} \alpha_{1}^{2} \alpha_{2}^{2} \beta^{\ast} \delta^{\ast}}{2} \\
 & + & \left. \frac{l_{1}^{\star} l_{6} \left( \left|\alpha_{1}\right|^{2} + 1 \right) \alpha \gamma \delta^{\ast}}{2} + \frac{l_{1}^{\star} l_{7} \left|\alpha_{2}\right|^{2} \alpha_{1}^{\ast 2} \alpha_{2}^{\ast 2} \beta \delta}{2} \right\} + {\rm c.c.} ]
\end{array}
\end{equation}

{   While investigating quantum entanglement, we derive the formula
for intermodal entanglement across various coupled modes 
using HZ-1 criteria (cf. Eq. (\ref{HZ-1=000020criteria})) }{ }

{  
\begin{equation}
\begin{array}{lcl}
E_{a_{1}b} & = & \left|g_{2}\right|^{2}\left(\left|\alpha_{1}\right|^{4}\left|\alpha_{2}\right|^{2}+\left|\beta\right|^{4}\left|\gamma\right|^{2}-\left|\alpha_{1}\right|^{2}\left|\beta\right|^{2}\left(\left|\alpha_{2}\right|^{2}+\left|\gamma\right|^{2}\right)\right)\\
 & + & \left|g_{3}\right|^{2}\left(\left|\alpha_{2}\right|^{2}+\left|\gamma\right|^{2}\right)\left|\beta\right|^{2}\left|\delta\right|^{2}+\left|g_{4}\right|^{2}\left|\alpha\right|^{2}\left|\beta\right|^{2}\\
 & + & \left[\left\{ -g_{1}g_{5}^{*}\left|\alpha_{1}\right|^{2}\beta^{\ast}\gamma^{\ast2}\delta-g_{1}g_{7}^{*}\left|\alpha_{1}\right|^{2}\alpha\beta^{*}\gamma^{*}+g_{2}g_{3}^{*}\left|\beta\right|^{2}\beta\gamma^{2}\delta^{*}\right.\right.\\
 & + & \left.\left.g_{2}g_{4}^{\ast}\left|\beta\right|^{2}\alpha^{*}\beta\gamma+g_{3}g_{4}^{*}\left|\beta\right|^{2}\alpha^{*}\gamma^{*}\delta\right\} +{\rm c.c.}\right]
\end{array}
\end{equation}
}{ }

\begin{equation}
\begin{aligned}
E_{a_{1}c} &= \left|g_{2}\right|^{2} \left( \left|\alpha_{1}\right|^{2} - \left|\gamma\right|^{2} \right) \left( \left|\alpha_{1}\right|^{2} \left|\alpha_{2}\right|^{2} - \left|\beta\right|^{2} \left|\gamma\right|^{2} \right) \\
& \quad + \left|g_{3}\right|^{2} \Bigg\{ 3 \left|\alpha_{1}\right|^{2} \left( \left|\alpha_{2}\right|^{2} + 1 \right) \left|\delta\right|^{2} + \left( \left|\alpha_{1}\right|^{2} + \left|\alpha_{2}\right|^{2} + 1 \right) \left|\gamma\right|^{2} \left|\delta\right|^{2} \\
& \quad + \left|\alpha_{1}\right|^{4} \left|\delta\right|^{2} + \left|\alpha_{2}\right|^{2} \left|\delta\right|^{2} - \left|\alpha_{1}\right|^{2} \left|\alpha_{2}\right|^{2} \left|\gamma\right|^{2} \Bigg\}  + \left|g_{4}\right|^{2} \left|\alpha\right|^{2} \left|\gamma\right|^{2} \\
& \quad + \Bigg[ \Bigg\{ g_{1} g_{3}^{*} \alpha_{1} \alpha_{2} \gamma \delta^{*} + 2 g_{1} g_{6}^{*} \left|\alpha_{1}\right|^{2} \beta \gamma^{2} \delta^{*} \\
& \quad + g_{2} g_{3}^{*} \left( - \left|\alpha_{1}\right|^{2} + \left|\alpha_{2}\right|^{2} + \left|\gamma\right|^{2} + 1 \right) \beta \gamma^{2} \delta^{*}  - g_{1} g_{7}^{*} \left|\alpha_{1}\right|^{2} \alpha \beta^{*} \gamma^{*} + g_{2}^{*} g_{4} \left|\gamma\right|^{2} \alpha \beta^{*} \gamma^{*} \\
& \quad + g_{1} g_{9}^{*} \left|\alpha_{1}\right|^{2} \alpha \gamma \delta^{*}  + g_{3}^{*} g_{4} \left( \left|\alpha_{2}\right|^{2} + \left|\gamma\right|^{2} + 1 \right) \alpha \gamma \delta^{*} - k_{2} k_{3}^{*} \alpha_{1}^{2} \alpha_{2}^{2} \beta^{*} \delta^{*} \Bigg\} + \text{{\rm c.c.}.} \Bigg]
\end{aligned}
\end{equation}

\begin{equation}
\begin{aligned}
E_{a_{1}d} = & \left|g_{2}\right|^{2} \left|\beta\right|^{2} \left|\gamma\right|^{2} \left|\delta\right|^{2} + \left|g_{3}\right|^{2} \left( \left|\delta\right|^{2} + 1 \right) \left\{ \left( \left|\alpha_{2}\right|^{2} + \left|\gamma\right|^{2} + 1 \right) \left|\delta\right|^{2} - \left|\alpha_{2}\right|^{2} \left|\gamma\right|^{2} \right\} \\
& + \left|g_{4}\right|^{2} \left|\alpha\right|^{2} \left|\delta\right|^{2} - \left|l_{2}\right|^{2} \left\{ \left|\alpha_{1}\right|^{2} \left( \left|\alpha_{2}\right|^{2} + \left|\gamma\right|^{2} + 1 \right) \left|\delta\right|^{2} + \left( \left|\alpha_{2}\right|^{2} + \left|\gamma\right|^{2} + 1 \right) \left|\delta\right|^{2} \right. \\
& - \left. \left|\alpha_{2}\right|^{2} \left|\gamma\right|^{2} \left( \left|\delta\right|^{2} + 1 \right) \right\} + \\
& + \left[ -g_{1} g_{6}^{*} \left|\alpha_{1}\right|^{2} \beta \gamma^{2} \delta^{*} - g_{1} g_{9}^{*} \left|\alpha_{1}\right|^{2} \alpha \gamma \delta^{*} + g_{2} g_{3}^{*} \left|\delta\right|^{2} \beta \gamma^{2} \delta^{*} \right. \\
& + \left. g_{2} g_{4}^{*} \alpha^{*} \beta \gamma \left|\delta\right|^{2} + g_{3} g_{4}^{*} \left|\delta\right|^{2} \alpha^{*} \gamma^{*} \delta \right] + {\rm c.c.}.
\end{aligned}
\end{equation}

{  
\begin{equation}
\begin{array}{lcl}
E_{bc} & = & \left|j_{2}\right|^{2}\left\{ \left|\alpha_{1}\right|^{2}\left|\alpha_{2}\right|^{2}\left(3\left|\beta\right|^{2}+3\left|\gamma\right|^{2}+1\right)-\left|\beta\right|^{2}\left|\gamma\right|^{2}\left(\left|\alpha_{1}\right|^{2}+\left|\alpha_{2}\right|^{2}+1\right)\right\} \\
 & + & \left[\left\{ j_{1}j_{2}^{\ast}\alpha_{1}^{\ast}\alpha_{2}^{\ast}\beta\gamma+k_{2}k_{3}^{\ast}\alpha_{1}^{2}\alpha_{2}^{2}\beta^{*}\delta^{*}+j_{1}j_{6}^{\ast}\left(\left|\alpha_{1}\right|^{2}+\left|\alpha_{2}\right|^{2}+1\right)\alpha^{*}\beta\gamma\right.\right.\\
 & + & \left.\left.2j_{1}j_{4}^{\ast}\left(\left|\alpha_{1}\right|^{2}+\left|\alpha_{2}\right|^{2}+1\right)\beta\gamma^{2}\delta^{\ast}\right\} +{\rm c.c.}\right]
\end{array}
\end{equation}
}{ }

{  
\begin{equation}
\begin{array}{lcl}
E_{bd} & = & \left|j_{2}\right|^{2}\left|\alpha_{1}\right|^{2}\left|\alpha_{2}\right|^{2}\left|\delta\right|^{2}+\left[l_{1}^{\ast}l_{5}\alpha_{1}^{2}\alpha_{2}^{2}\beta^{\ast}\delta^{\ast}+{\rm c.c.}\right]\end{array}
\end{equation}
}{ }

{  
\begin{equation}
\begin{array}{lcl}
E_{cd} & = & \left|k_{2}\right|^{2}\left|\alpha_{1}\right|^{2}\left|\alpha_{2}\right|^{2}\left|\delta\right|^{2}+\left|k_{3}\right|^{2}\left(\left|\alpha_{1}\right|^{2}+\left|\alpha_{2}\right|^{2}+1\right)\left|\delta\right|^{2}\left(\left|\gamma\right|^{2}+\left|\delta\right|^{2}+2\right)\end{array}
\end{equation}
}{ }

{   In the similar manner, we obtain the expression for the various
combined modes using HZ-2 criteria (\ref{HZ-2=000020criteria}). These
are }{ }

\begin{equation}
\begin{array}{lcl}
E_{a_{1}b}^{'} & = & \left|g_{2}\right|^{2} \left( \left|\beta\right|^{4} \left|\gamma\right|^{2} + \left|\alpha_{1}\right|^{4} \left|\alpha_{2}\right|^{2} + \left|\alpha_{1}\right|^{2} \left|\alpha_{2}\right|^{2} \left|\beta\right|^{2} + \left|\alpha_{1}\right|^{2} \left|\beta\right|^{2} \left|\gamma\right|^{2} \right) \\
& + & \left|g_{3}\right|^{2} \left( \left|\alpha_{2}\right|^{2} + \left|\gamma\right|^{2} + 1 \right) \left|\beta\right|^{2} \left|\delta\right|^{2} + \left|g_{4}\right|^{2} \left|\alpha\right|^{2} \left|\beta\right|^{2} \\
& + & \left[ \left\{ g_{2} g_{3}^{*} \beta \gamma^{2} \delta^{*} \left|\beta\right|^{2} + g_{2} g_{4}^{*} \alpha^{*} \beta \gamma \left|\beta\right|^{2} \right. \right. \\
& + & \left. \left. g_{3} g_{4}^{*} \alpha^{*} \gamma^{*} \delta \left|\beta\right|^{2} - j_{1}^{*} j_{4} \left|\alpha_{1}\right|^{2} \beta^{*} \gamma^{*2} \delta - j_{1} j_{6}^{*} \left|\alpha_{1}\right|^{2} \alpha^{*} \beta \gamma \right\} + {\rm c.c.}. \right]
\end{array}
\end{equation}

\begin{equation}
\begin{array}{lcl}
E_{a_{1}c}^{'} & = & \left|g_{2}\right|^{2} \left( \left|\beta\right|^{2} \left|\gamma\right|^{4} + \left|\alpha_{1}\right|^{4} \left|\alpha_{2}\right|^{2} + \left|\alpha_{1}\right|^{2} \left|\alpha_{2}\right|^{2} \left|\gamma\right|^{2} - \left|\alpha_{1}\right|^{2} \left|\beta\right|^{2} \left|\gamma\right|^{2} \right) \\
& + & \left|g_{3}\right|^{2} \left\{ \left( \left|\alpha_{2}\right|^{2} + \left|\gamma\right|^{2} + 1 \right) \left|\gamma\right|^{2} \left|\delta\right|^{2} - 2 \left|\alpha_{2}\right|^{2} \left|\gamma\right|^{2} \left|\delta\right|^{2} + \left|\alpha_{1}\right|^{4} \left|\delta\right|^{2} \right. \\
& - & \left. \left|\alpha_{1}\right|^{2} \left|\alpha_{2}\right|^{2} \left|\delta\right|^{2} + \left|\alpha_{1}\right|^{2} \left|\delta\right|^{2} - \left|\alpha_{2}\right|^{2} \left|\delta\right|^{2} \right\} \\
& + & \left|g_{4}\right|^{2} \left|\alpha\right|^{2} \left|\gamma\right|^{2} + \left[ \left\{ -k_{1} k_{14}^{*} \left|\alpha_{1}\right|^{2} \beta \gamma^{2} \delta^{*} \right. \right. \\
& - & k_{1} k_{4}^{*} \left|\alpha_{1}\right|^{2} \alpha^{*} \beta \gamma - k_{1}^{*} k_{6} \left|\alpha_{1}\right|^{2} \alpha^{*} \gamma^{*} \delta + g_{2} g_{3}^{*} \left( \left|\gamma\right|^{2} - \left|\alpha_{2}\right|^{2} \right) \beta \gamma^{2} \delta^{*} \\
& + & g_{2} g_{4}^{*} \left|\gamma\right|^{2} \alpha^{*} \beta \gamma + \left. \left. g_{3} g_{4}^{*} \left( \left|\gamma\right|^{2} - \left|\alpha_{2}\right|^{2} \right) \alpha^{*} \gamma^{*} \delta - k_{2} k_{3}^{*} \alpha_{1}^{2} \alpha_{2}^{2} \beta^{*} \delta^{*} \right\} + {\rm c.c.}. \right]
\end{array}
\end{equation}

\begin{equation}
\begin{array}{lcl}
E_{a_{1}d}^{'} & = & \left|g_{2}\right|^{2} \left( \left|\beta\right|^{2} \left|\gamma\right|^{2} + \left|\alpha_{1}\right|^{2} \left|\gamma\right|^{2} + \left|\alpha_{1}\right|^{2} \left|\alpha_{2}\right|^{2} + \left|\alpha_{1}\right|^{2} \right) \left|\delta\right|^{2} \\
& + & \left|g_{3}\right|^{2} \left( \left|\alpha_{2}\right|^{2} + \left|\gamma\right|^{2} + 1 \right) \left|\delta\right|^{4} + \left|g_{4}\right|^{2} \left|\alpha\right|^{2} \left|\delta\right|^{2} \\
& + & \left[ \left\{ g_{2} g_{3}^{*} \beta \gamma^{2} \delta^{*} \left|\delta\right|^{2} + g_{2} g_{4}^{*} \alpha^{*} \beta \gamma \left|\delta\right|^{2} + g_{3} g_{4}^{*} \alpha^{*} \gamma^{*} \delta \left|\delta\right|^{2} \right. \right. \\
& - & \left. \left. l_{1}^{\ast} l_{3} \left|\alpha_{1}\right|^{2} \beta \gamma^{2} \delta^{\ast} - l_{1} l_{6}^{\ast} \left|\alpha_{1}\right|^{2} \alpha^{*} \gamma^{*} \delta \right\} + {\rm c.c.}. \right]
\end{array}
\end{equation}

{  
\begin{equation}
\begin{array}{lcl}
E_{bc}^{'} & = & -\left|j_{2}\right|^{2}\left|\alpha_{1}\right|^{2}\left|\alpha_{2}\right|^{2}\left(\left|\beta\right|^{2}+\left|\gamma\right|^{2}+1\right)+\left|k_{3}\right|^{2}\left(\left|\alpha_{1}\right|^{2}+\left|\alpha_{2}\right|^{2}+1\right)\left|\beta\right|^{2}\left|\delta\right|^{2}\\
 & - & \left[\left\{ k_{1}k_{2}^{\ast}\alpha_{1}^{\ast}\alpha_{2}^{\ast}\beta\gamma+k_{1}k_{4}^{\ast}\left(\left|\alpha_{1}\right|^{2}+\left|\alpha_{2}\right|^{2}+1\right)\alpha^{*}\beta\gamma+k_{2}k_{3}^{\ast}\alpha_{1}^{2}\alpha_{2}^{2}\beta^{*}\delta^{*}\right.\right.\\
 & + & \left.\left.\left(j_{1}j_{2}^{\ast}k_{1}k_{3}^{\ast}+k_{1}k_{14}^{\ast}\right)\left(\left|\alpha_{1}\right|^{2}+\left|\alpha_{2}\right|^{2}+1\right)\beta\gamma^{2}\delta^{*}\right\} +{\rm c.c.}\right]
\end{array}
\end{equation}
}{ }

{  
\begin{equation}
\begin{array}{lcl}
E_{bd}^{'} & = & \left|j_{2}\right|^{2}\left|\alpha_{1}\right|^{2}\left|\alpha_{2}\right|^{2}\left|\delta\right|^{2}-\left[l_{1}^{\ast}l_{5}\alpha_{1}^{2}\alpha_{2}^{2}\beta^{\ast}\delta^{\ast}+{\rm c.c.}\right]\end{array}
\end{equation}
}{ }

{  
\begin{equation}
\begin{array}{lcl}
E_{cd}^{'} & = & \left|k_{2}\right|^{2}\left|\alpha_{1}\right|^{2}\left|\alpha_{2}\right|^{2}\left|\delta\right|^{2}+\left|k_{3}\right|^{2}\left(\left|\alpha_{1}\right|^{2}+\left|\alpha_{2}\right|^{2}+1\right)\left(\left|\gamma\right|^{2}+\left|\delta\right|^{2}\right)\left|\delta\right|^{2}\end{array}
\end{equation}
}{ }

Similarly, we obtain the expressions of the witnesses for boson antibunching in all the modes of hyper-Raman system as follows:
\begin{equation}
\begin{array}{lcl}
D_{a_{1}} & = & 2\left|g_{2}\right|^{2}\left|\alpha_{1}\right|^{2}\left|\beta\right|^{2}\left|\gamma\right|^{2} + 2\left|g_{3}\right|^{2}\left|\alpha_{1}\right|^{2}\left(\left|\alpha_{2}\right|^{2} + \left|\gamma\right|^{2} + 1\right)\left|\delta\right|^{2} \\
& + & 2\left|g_{4}\right|^{2}\left|\alpha\right|^{2}\left|\alpha_{1}\right|^{2} + \left[ \left\{ g_{1}^{*}g_{2} g_{1}^{*} g_{3} \alpha_{1}^{\ast 2} \alpha_{2}^{\ast 2} \beta \delta + 2g_{2} g_{3}^{*} \left|\alpha_{1}\right|^{2} \beta \gamma^{2} \delta^{*} \right. \right. \\
& + & 2g_{2} g_{4}^{*} \alpha^{\ast} \left|\alpha_{1}\right|^{2} \beta \gamma + 2g_{3} g_{4}^{*} \alpha^{\ast} \left|\alpha_{1}\right|^{2} \gamma^{\ast} \delta \left\} + {\rm c.c.}. \right]
\end{array}
\end{equation}

{  
\begin{equation}
\begin{array}{lcl}
D_{b} & = & 2\left|j_{2}\right|^{2}\left|\alpha_{1}\right|^{2}\left|\alpha_{2}\right|^{2}\left|\beta\right|^{2}\end{array}
\end{equation}

\begin{equation}
\begin{array}{lcl}
D_{c} & = & 2\left|k_{2}\right|^{2}\left|\alpha_{1}\right|^{2}\left|\alpha_{2}\right|^{2}\left(\left|\beta\right|^{2}+1\right)\left|\gamma\right|^{2}+2\left|k_{3}\right|^{2}\left(\left|\alpha_{1}\right|^{2}+1\right)\left(\left|\alpha_{2}\right|^{2}+1\right)\left|\gamma\right|^{2}\left|\delta\right|^{2}\\
 & + & \left[k_{1}^{*}k_{2}k_{1}^{*}k_{3}\left(\left|\alpha_{1}\right|^{2}+\left|\alpha_{2}\right|^{2}+1\right)\beta^{\ast}\gamma^{\ast2}\delta+{\rm c.c.}\right]
\end{array}
\end{equation}
\begin{equation}
\begin{array}{lcl}
D_{d} & = & 2\left|l_{2}\right|^{2}\left|\alpha_{1}\right|^{2}\left|\alpha_{2}\right|^{2}\left|\gamma\right|^{2}\left|\delta\right|^{2}\end{array}
\end{equation}

\end{document}